\documentclass[aps,english,preprint,longbibliography]{revtex4-1}
\usepackage{amsmath}
\usepackage{amssymb}
\usepackage{subfigure}
\usepackage[pdftex]{graphicx}
\usepackage[pdftex,bookmarks=true,bookmarksopen,bookmarksnumbered,
                colorlinks,
                linkcolor=blue,
                citecolor=blue]{hyperref}
\usepackage{dcolumn}
\usepackage{bm}
\usepackage{braket,amssymb,bbm}
\usepackage{upgreek}
\usepackage{color}
\usepackage{ulem}
\usepackage{placeins}

\begin{document}


\title{Magnetic resonance with squeezed microwaves}

\author{A. Bienfait$^{1}$, P. Campagne-Ibarcq$^{1}$, A. Holm-Kiilerich$^{2}$, X. Zhou$^{1,3}$, S. Probst$^{1}$, J.J. Pla$^{4}$, T. Schenkel$^{5}$, D. Vion$^{1}$, D. Esteve$^{1}$, J.J.L. Morton$^{6}$, K. Moelmer$^{2}$, and P. Bertet$^{1}$}

\affiliation{$^{1}$Quantronics group, SPEC, CEA, CNRS, Universit\'e Paris-Saclay, CEA Saclay 91191 Gif-sur-Yvette Cedex, France}

\affiliation{$^{2}$Department of Physics and Astronomy, Aarhus University, Ny Munkegade 120, DK-8000 Aarhus C, Denmark}
	
\affiliation{$^{3}$Institute of Electronics Microelectronics and Nanotechnology, CNRS UMR 8520, ISEN Department, Avenue Poincar\'e, CS 60069, 59652 Villeneuve d'Ascq Cedex, France}

\affiliation{$^{4}$School of Electrical Engineering and Telecommunications, University of New South Wales, Anzac Parade, Sydney, NSW 2052, Australia}

\affiliation{$^{5}$Accelerator Technology and Applied Physics Division, Lawrence Berkeley National Laboratory, Berkeley,
California 94720, USA}

\affiliation{$^{6}$ London Centre for Nanotechnology, University College London, London WC1H 0AH, United Kingdom}

\date{\today}

\newcommand{\pb}{\textcolor{black}}
\newcommand{\ay}{\textcolor{black}}
\newcommand{\ak}{\textcolor{black}}

\begin{abstract}
\pb{Vacuum fluctuations of the electromagnetic field set a fundamental limit to the sensitivity of a variety of measurements, including magnetic resonance spectroscopy. We report the use of squeezed microwave fields, which are engineered quantum states of light for which fluctuations in one field quadrature are reduced below the vacuum level, to enhance the detection sensitivity of an ensemble of electronic spins at millikelvin temperatures.} By shining a squeezed vacuum state on the input port of a microwave resonator containing the spins, we obtain a $1.2$\,dB noise reduction at the spectrometer output compared to the case of a vacuum input. This result constitutes a proof of principle of the application of quantum metrology to magnetic resonance spectroscopy.
\end{abstract}

\maketitle

\section{Introduction}

The detection and characterisation of electron spins in a sample by magnetic resonance spectroscopy~\cite{SchweigerEPR(2001)} has numerous applications in materials science, chemistry, and quantum information processing. Pulsed magnetic resonance detection proceeds by detecting weak microwave signals emitted by spins resonant with a cavity in which the sample is embedded. The noise present in these signals determines the spectrometer sensitivity and is ultimately limited by the fluctuations in the microwave field at the cavity output. The thermal contribution to these fluctuations can be removed by lowering the temperature $T$ of the sample and cavity such that $k_B T \ll \hbar \omega_s $, where $\omega_s$ is the spin resonance frequency and $k_B$ is Boltzmann's constant~\cite{bienfait2015reaching}. However, even at these cryogenic temperatures, quantum fluctuations of the electromagnetic field remain and pose a fundamental limitation to the achievable sensitivity.

\begin{figure}[h!]
\centering
\includegraphics[width=120mm]{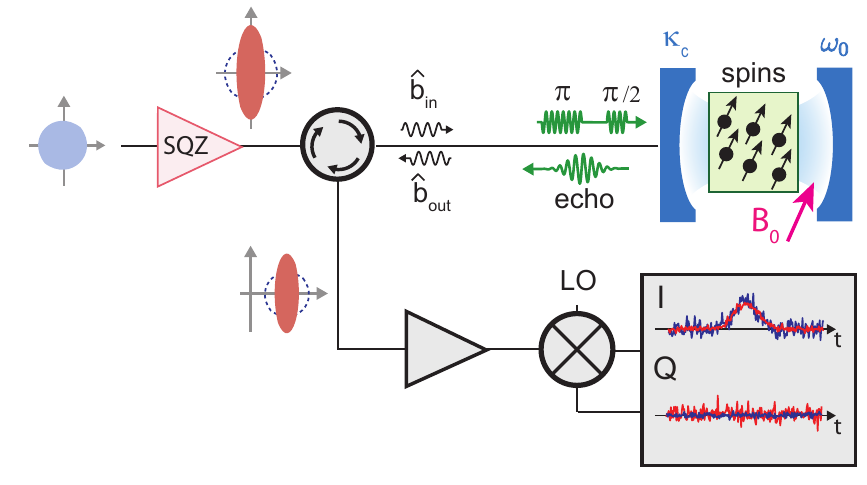}
\caption{Principle of squeezing-enhanced pulsed magnetic resonance. A squeezed vacuum state is incident on an ESR cavity of frequency $\omega_0$. The cavity contains the spins to be detected, which are tuned into resonance at $\omega_0$ by a dc magnetic field $B_0$. A Hahn echo microwave pulse sequence ($\pi / 2 - \tau - \pi - \tau$) is applied to the spins, leading to the emission of an echo in the detection waveguide on the $X$ quadrature. This echo is noiselessly amplified along $X$ before its homodyne demodulation with a local oscillator (L.O.) phase such that $I(t)$ is proportional to $X(t)$. The traces in the bottom right grey box, which are not real data, depict schematically the expected difference between SQZ off (blue) and SQZ on (red) output quadrature signals when the squeezed quadrature is aligned along the echo emission quadrature $X$; the signal-to-noise ratio is improved on the $I$ quadrature which contains the entire echo signal. \label{fig1} }
\end{figure}

Field fluctuations are governed by Heisenberg's uncertainty principle, which states that $\delta X^2 \delta Y^2 \geq 1/16$. In this expression, $\hat{X}$ and $\hat{Y}$ are the two quadrature operators of the field in dimensionless units, normalized such that $\langle \hat{X}^2 \rangle + \langle \hat{Y}^2 \rangle = N + 1/2$, $N$ being the average photon number in the field mode of interest. When the field is in a coherent state, which is the case for the echo signals emitted by the spins, $\delta X^2 = \delta Y^2 = 1/4$, as in the vacuum state. It is possible, however, to engineer so-called squeezed states in which the variance in one quadrature (called the squeezed quadrature) is reduced below $1/4$, at the expense of an increase in variance in the other quadrature as required by Heisenberg's inequality (see Fig.~\ref{fig1}). Most experiments demonstrating the production~\cite{Slusher.PhysRevLett.55.2409(1985),Wu.PhysRevLett.57.2520} and use of squeezed states have been performed in the optical domain. Squeezed optical states have been used to enhance the sensitivity of interferometric measurements ~\cite{PhysRevD.23.1693(1981),Grangier.PhysRevLett.59.2153(1987),Xiao.PhysRevLett.59.278(1987)} with applications in gravitational wave detection~\cite{ligo2011gravitational,ligo2013enhanced}, atomic absorption spectroscopy~\cite{Polzik.PhysRevLett.68.3020(1992)}, imaging~\cite{treps2003quantum}, atom-based magnetometry~\cite{Lucivero.PhysRevA.93.053802(2016)}, and of particle tracking in biological systems~\cite{taylor2013biological}. 

At microwave frequencies, the need to operate at cryogenic temperatures and lack of applications limited the interest in squeezed states to pioneering proof-of-principle demonstrations~\cite{movshovich_observation_1990} until the recent advent of quantum information processing with superconducting circuits~\cite{Devoret1169} which requires the control and measurement of microwave fields at the quantum level. This triggered the development of practical Josephson Parametric Amplifier (JPA) devices~\cite{Castellanos-Beltran.ApplPhysLett.91.083509(2007),bergeal_phase-preserving_2010,Zhou.PhysRevB.89.214517(2014),macklin2015near} and of follow-up amplifier chains such that the output noise is dominated by quantum fluctuations~\cite{mallet_quantum_2011}. Microwave squeezed states~\cite{mallet_quantum_2011} can then provide a sizeable noise reduction, thus improving measurement sensitivity for qubit state readout~\cite{Didier.PhysRevLett.115.093604(2015),Didier.PhysRevLett.115.203601(2015),eddins2017stroboscopic,govia2017enhanced} and nanomechanical resonator motion detection~\cite{Clark.NatPhys.2016}. They have also been investigated for their effect on the dynamics of quantum systems, such as two-level atoms~\cite{Gardiner.PhysRevLett.56.1917(1985),murch2013reduction,toyli2016resonance} or mechanical oscillators~\cite{clark2017sideband}. Here we propose and demonstrate a novel application of quantum squeezing at microwave frequencies to magnetic resonance spectroscopy for improving the detection sensitivity of a small ensemble of electronic spins.

Our scheme is depicted in Fig.~\ref{fig1}. A single-port lossless resonator of frequency $\omega_0$ containing the sample is coupled with a rate $\kappa_C$ to the measurement line that supports incoming ($\hat{b}_{\mathrm{in}}$) and outgoing ($\hat{b}_{\mathrm{out}}$) field modes. A dc magnetic field $B_0$ is applied to tune the spin frequency into resonance with the cavity. A Hahn echo sequence~\cite{Hahn.PhysRev.80.580(1950)} consisting of a $\pi / 2$ pulse at time $t=0$ followed by a $\pi$ pulse at $t=\tau$ leads to rephasing of the spins at $t=2\tau$ and, whenever $\omega_s \approx \omega_0$, to the emission of an echo on one field quadrature (that we take here to be $X$) in the output measurement line. This echo is then amplified noiselessly with a quantum-limited amplifier tuned to the $X$ quadrature~\cite{bienfait2015reaching}, and detected by coherently mixing it with a local oscillator. Throughout the article we will consider that the local oscillator phase is adjusted such that the echo signal at the measurement chain output (i.e. after propagation and further amplification) is entirely on one of the two quadratures that we will call $I(t)$, the other being denoted $Q(t)$. The noise accompanying the echo arises from the (amplified) fluctuations in $\hat{b}_{\mathrm{in}}$, the input field reflected by the cavity into the output mode; it reaches the quantum limit if $\hat{b}_{\mathrm{in}}$ is in the vacuum state. If $\hat{b}_{\mathrm{in}}$ is instead prepared in a squeezed vacuum state with its squeezed quadrature coinciding with the $X$ quadrature on which the echo is emitted, the noise in $I(t)$ may be below the quantum limit (see Fig.~\ref{fig1}), whereas the noise on the other quadrature (which bears no spin signal) is correspondingly increased.

\pb{Such a noise reduction at constant echo amplitude implies that the sensitivity of spin-echo detection, defined as the minimum number of spins that can be detected with unit signal-to-noise ratio in a given measurement time, can be improved beyond the limit imposed by vacuum fluctuations of the microwave field simply by sending a squeezed state onto the cavity input. For a given sample, the same signal-to-noise ratio can thus be reached in a shorter measurement time using squeezed states. This may have practical applications in magnetic resonance, in particular for samples with very low spin concentrations, or even containing only one spin~\cite{Haikka.PhysRevA.95.022306}. It also raises an interesting fundamental question about the ultimate limits on the signal-to-noise ratio achievable in spin-echo detection, given that the electromagnetic noise contribution can be fully suppressed by combining quantum squeezing and noiseless amplification.}

\pb{The purpose of this article is twofold. First, in Sec.~\ref{sec:theory} we present a theoretical analysis of the experiment proposed above. The results confirm that, in the limit where the coupling of the spin ensemble to the resonator is weak, squeezed state injection does lead to reduced noise in the echo signal. Second, we present an experimental implementation of this proposal. In Sec.~\ref{sec:squeezing} we characterize squeezed vacuum radiation generated by a Josephson Parametric Amplifier (JPA) at millikelvin temperatures, and we analyse the physical effects that limit the noise reduction to $1.2$\,dB. In Sec.~\ref{sec:ESRWithSqueezing}, we demonstrate that applying the squeezed microwave source to the ESR spectrometer increases its sensitivity by the same amount. In Sec.~\ref{sec:DiscussionConclusion}, we conclude with a discussion of the practical and fundamental interest and limitations of our scheme.}

\section{Spin-echo emission in squeezed vacuum state : theory}
\label{sec:theory}

Our physical system, illustrated in Fig.~\ref{fig1}b, consists of a cavity mode coupled resonantly to $N_{\mathrm{spins}}$ spins and to input and output microwave fields.  As we are interested in both the mean amplitude and the quantum fluctuations in the output signal, we describe the whole system quantum mechanically. Inhomogeneous broadening and spatial variations of the spins within the ESR cavity lead to different transition frequencies $\omega_j$ and coupling strengths $g_j$ of the individual spins to the cavity mode. We will assume here that the spins are close to resonance with the cavity, and that their mean frequency $\omega_s$ is equal to the cavity frequency $\omega_0$. In a frame rotating at $\omega_0$, the total Hamiltonian of the spins and the resonator mode is
\begin{align}
\hat{H} = \hbar\sum_j\left[g_j\left(\hat{\sigma}_-^{(j)}\hat{a}^\dagger+\hat{\sigma}_+^{(j)}\hat{a}\right)+\frac{\Delta_j}{2}\hat{\sigma}_z^{(j)}\right],
\end{align}
where $\Delta_j = \omega_j-\omega_0$ denotes the detuning of the j$^{th}$ spin from the cavity resonance frequency, $\hat{a}$ and $\hat{a}^\dagger$ denote field annihilation and creation operators, and $\hat{\sigma}_z^{(j)}$, $\hat{\sigma}_{-(+)}^{(j)}$ are Pauli operators describing the spin degrees of freedom.

The quantum-optical input-output formalism~\cite{Gardiner.PhysRevA.31.3761(1985)} yields the following Heisenberg equation for the cavity field operator:
\begin{equation} \label{eq:fieldL}
\dot{\hat{a}}= -i\sum_j g_j\hat{\sigma}_-^{(j)} -\frac{\kappa}{2} \hat{a} + \sqrt{\kappa_L}\hat{b}_{\mathrm{loss}}(t)+ \sqrt{\kappa_C}\hat{b}_{\mathrm{in}}(t),
\end{equation}
where $\kappa=\kappa_C+\kappa_L$ is the total cavity damping rate with contributions $\kappa_C$ due to the out-coupling and $\kappa_L$ due to internal cavity losses.
The last two terms in Eq.~\ref{eq:fieldL} describe inputs from bath modes: $\hat{b}_{\mathrm{loss}}(t)$ associated with the internal cavity losses and $\hat{b}_{\mathrm{in}}(t)$ associated with the quantized radiation field incident on the cavity.

Rather than solving the complete excitation dynamics of the spins, we will assume that ideal $\pi/2$ and $\pi$ control pulses have been applied to the spins at times $t = -  \tau$ and $t=0$, respectively, preparing a state where the spin excited states have acquired phases $\exp(i\Delta_j \tau)$ with respect to the spin ground states in a frame rotating at $\omega_s$. As the spins precess at different frequencies $\Delta_j$, they come back in phase at the later time $t=\tau$, and we shall analyze their coupling to the quantized field during the rephasing of the spins that leads to the emission of an echo of duration $T_{\mathrm{E}}$, \ay{set by the spin spectral linewidth and the duration of the $\pi/2$ and $\pi$ control pulses.}

To this end we apply the so-called Holstein-Primakoff approximation~\cite{Holstein.PhysRev.58.1098(1940)}, which assumes oscillator-like commutator relations $[\hat{\sigma}_-^{(j)},\hat{\sigma}_+^{(k)}] = \delta_{jk}$ for the spin lowering operators $\hat{\sigma}_-^{(j)}$\ay{, and we treat each spin as an oscillator prepared in a coherent state of complex amplitude $\alpha \exp(i\Delta_j \tau)$ at $t=0$. The precession about the spin z-axis due to the inhomogeneous distribution of spin excitation energies is equivalent to the rotation of the complex oscillator amplitude, while the oscillator approximation assigns a constant damping rate to the collective transverse spin components and a linear coupling of the spin and field oscillator amplitudes rather than the non-linear, excitation-dependent one. Since the decay of the transverse spin components is very limited during the timescale of our protocol, describing it with a constant effective rate constant and assuming a linear oscillator-like coupling to the field is a good approximation.} We solve the coupled dynamics of the field mode and the spin ensemble, and we hence need the Heisenberg equation of motion for the spin lowering operator, which incorporates the coherent state initial condition as a delta-function excitation pulse at $t=0$,
\begin{equation} \label{eq:spins}
\dot{\hat{\sigma}}_j = -(\gamma+i\Delta_j)\hat{\sigma}_-^{(j)} -i g_j \hat{a} + \alpha \mathrm{e}^{i\Delta_j \tau}\delta(t) + \sqrt{2\gamma}\hat{F}_j(t).
\end{equation}
The relaxation rate $\gamma$ represents spin decoherence, and is accompanied by quantum Langevin noise sources $\hat{F}_j(t)$ with non-vanishing commutators $[\hat{F}_i(t),\hat{F}_j^\dagger(t^{\prime})]= \delta(t-t^{\prime})\delta_{ij}$.

An analysis of the beam-splitter like coupling of the incident, resonator and outgoing fields~\cite{Gardiner.PhysRevA.31.3761(1985)} yields the input-output relation,
\begin{equation} \label{eq:input_output}
\hat{b}_{\mathrm{out}}(t)= \sqrt{\kappa_C}\hat{a}(t) - \hat{b}_{\mathrm{in}}(t).
\end{equation}

Applying Fourier transforms, and solving the resulting algebraic set of equations for the coupled spin and field operators leads to the compact and general form of the output field operator,
 \begin{align} \label{eq:out}
 \begin{split}
\tilde{b}_{\mathrm{out}}(\omega)&= -\frac{iq(\omega)}{\sqrt{2\pi}} + t(\omega)\tilde{f}_\mathrm{spin}(\omega) +l(\omega) \tilde{b}_{\mathrm{loss}}(\omega) + r(\omega) \tilde{b}_{\mathrm{in}}(\omega),
\end{split}
\end{align}
where 
\begin{align}
q(\omega) = \frac{2\sqrt{\kappa_C} A(\omega)}{\kappa [1+ C(\omega)]-2i\omega },
\end{align}
\begin{align}\label{eq:transm}
t(\omega) = \frac{2\sqrt{\kappa_C\kappa\mathrm{Re}\left[C(\omega)\right]}}{\kappa[1+C(\omega)]-2i\omega},
\end{align}
\begin{align}\label{eq:l}
l(\omega) = \frac{2\sqrt{\kappa_L\kappa\mathrm{Re}\left[C(\omega)\right]}}{\kappa[1+C(\omega)]-2i\omega}
\end{align}
and
 \begin{equation} \label{eq:refl}
r(\omega) = \frac{\kappa_C-\kappa_L-(\kappa_C+\kappa_L)C(\omega)+2i\omega}{\kappa[1+C(\omega)]-2i\omega},
\end{equation}
are frequency-dependent complex coefficients describing respectively the mean field emitted by the spins and Langevin noise operator terms associated with the spins, the resonator internal loss and reflection of the microwave field on the cavity.

In Eqs.~\ref{eq:out}-\ref{eq:l}, the distribution of spin detunings and coupling strengths are incorporated in the frequency-dependent ensemble cooperativity,
\begin{equation}\label{eq:C}
C(\omega) = \sum_j \frac{2g_j ^2}{\kappa(\gamma+i\Delta_j-i\omega)}
\end{equation}
and the amplitude factor,
\begin{equation}
A(\omega) = \sum_j \frac{g_j\alpha e^{i\Delta_j \tau}}{\gamma+i\Delta_j-i\omega}.
\end{equation}

The noise operators $\tilde{f}_\mathrm{spin}(\omega)=\tilde{F}_{\mathrm{spin}}(\omega)/\sqrt{\kappa\mathrm{Re}\left[C(\omega)\right]}$, $\tilde{b}_{\mathrm{loss}}(\omega)$ and $\tilde{b}_{\mathrm{in}}(\omega)$ obey standard commutator relations, e.g., $[\tilde{f}_{\mathrm{spin}}(\omega),\tilde{f}_{\mathrm{spin}}^\dagger(\omega^\prime)] = \delta(\omega-\omega^\prime)$, and the condition $|r(\omega)|^2+|t(\omega)|^2+|l(\omega)|^2=1$ ensures the same commutator relation applies to the output field operators $\tilde{b}_{\mathrm{out}}(\omega)$. We refer to the Supplemental Material for details of the derivation of the general expressions and for analytical results in the special case of a Lorentzian detuning distribution uncorrelated with the coupling strengths.

We now turn to the definition of the modes on which the echo is emitted in order to define and estimate the measurement sensitivity. For the sake of simplicity we assume that the bandwidth $T_{\mathrm{E}}^{-1}$ of the spin-echo signal is narrower than the bandwidth of the squeezed radiation and of the resonator. The output signal mode is defined as $\hat{b}_{\mathrm{mode}}= (1/\sqrt{T_{\mathrm{E}}})\int_{\tau-T_{\mathrm{E}}/2}^{\tau+T_{\mathrm{E}}/2} \hat{b}_{\mathrm{out}}(t')\,dt'$, its $\hat{X}$ quadrature operator being $\hat{X}=\frac{1}{2i}(\hat{b}_{\mathrm{mode}}-\hat{b}^\dagger_{\mathrm{mode}})$. The normalization is chosen such that $\hat{b}_{\mathrm{mode}}^\dagger \hat{b}_{\mathrm{mode}}$ is the photon number (operator) in the mode. We similarly introduce $\hat{b}_{\mathrm{in}}= (1/\sqrt{T_{\mathrm{E}}})\int_{\tau-T_{\mathrm{E}}/2}^{\tau+T_{\mathrm{E}}/2} \hat{b}_{\mathrm{in}}(t')\,dt'$, $\hat{f}_{\mathrm{spin}}= (1/\sqrt{T_{\mathrm{E}}})\int_{\tau-T_{\mathrm{E}}/2}^{\tau+T_{\mathrm{E}}/2} \hat{f}_{\mathrm{spin}}(t')\,dt'$ and $\hat{b}_{\mathrm{loss}}= (1/\sqrt{T_{\mathrm{E}}})\int_{\tau-T_{\mathrm{E}}/2}^{\tau+T_{\mathrm{E}}/2} \hat{b}_{\mathrm{loss}}(t')\,dt'$, as well as their respective $\hat{X}_{\mathrm{in}},\hat{X}_{\mathrm{bath}},\hat{X}_{\mathrm{loss}}$ quadrature operators. The mean integrated amplitude of the spin-echo signal is given by the mean value of the $\hat{X}$ operator,
\begin{equation}\label{eq:betaMode}
\langle \hat{X} \rangle =  \frac{-i q(0)}{\sqrt{T_{\mathrm{E}}}}.
\end{equation}
From Eq.~\ref{eq:out} one obtains that its fluctuations are
\begin{align}\label{eq:noise}
\delta X^2 &= |r(0)|^2 \delta X_{\mathrm{in}}^2 +|l(0)|^2 \delta X_{\mathrm{bath}}^2 + |t(0)|^2 \delta X_{\mathrm{spin}}^2.
\end{align}

In a model where the spins are described as harmonic oscillators coupled via the $F_j$ to an effective zero temperature bath,  $\delta X_{\mathrm{spin}}^2=\frac{1}{4}[\braket{[\tilde{f}_\mathrm{spin},\tilde{f}^\dagger_\mathrm{spin}]}+2\braket{\tilde{f}^\dagger_\mathrm{spin}\tilde{f}_\mathrm{spin}}]$ is equal to $1/4$. A more realistic description of the spins, going beyond the Holstein Primakoff approximation and taking into account a non-zero effective temperature, would yield a larger value but still of order unity. Since $|t(0)|^2 = \frac{\kappa_C}{\kappa} (1-|r(0)|^2) \simeq 4\frac{\kappa_C}{\kappa} C(0)$, the contribution of the spin fluctuation to the total output noise scales as the ensemble cooperativity $C(0)$. 

In the limit where the ensemble cooperativity and the cavity losses are small ($C(0) \ll 1$ and $|l(0)|^2 \ll 1$), which is the case in our experiment as explained in the next sections, the dominant contribution to the output fluctuations therefore is the reflected input noise and Eq.~\ref{eq:noise} reduces to $\delta X \approx  \delta X_{\mathrm{in}}$. This input field originates from a squeezing source (SQZ in Fig.~\ref{fig1}b) that we assume to be ideal, generating a squeezed vacuum along the $X$ quadrature with a variance $\delta X_{\mathrm{sq}}^2$ at its output. Due to transmission losses between SQZ and the ESR cavity, modelled by an effective loss coefficient $\eta_{\rm loss}$, the squeezing properties are deteriorated and the variance in the input quadrature becomes:

\begin{equation}\label{eq:etaloss}
\delta X_{\mathrm{in}}^2 = (1 - \eta_{\rm loss})\delta X_{\mathrm{sq}}^2+\eta_{\rm loss}/4,
\end{equation}

characterized by the ratio of the squeezed quadrature variance to the vacuum fluctuations $\eta_S = \delta X_{\mathrm{in}}^2 / (1/4)$ called the squeezing factor. The signal-to-noise ratio of the spin-echo detection is given by $\langle \hat{X} \rangle / \delta X$, $\langle \hat{X} \rangle$ being independent of the input field fluctuations as seen from Eq.~\ref{eq:betaMode}. Our analysis therefore shows that in the limit where the spin ensemble is weakly coupled to the cavity and the cavity losses are negligible, applying a squeezed vacuum to the ESR resonator should improve the spin detection sensitivity by approximately a factor $\sqrt{\eta_S}$. 

\section{Experimental results: Squeezed state characterization}
\label{sec:squeezing}

\begin{figure}[!htbp]
\includegraphics[width=85mm]{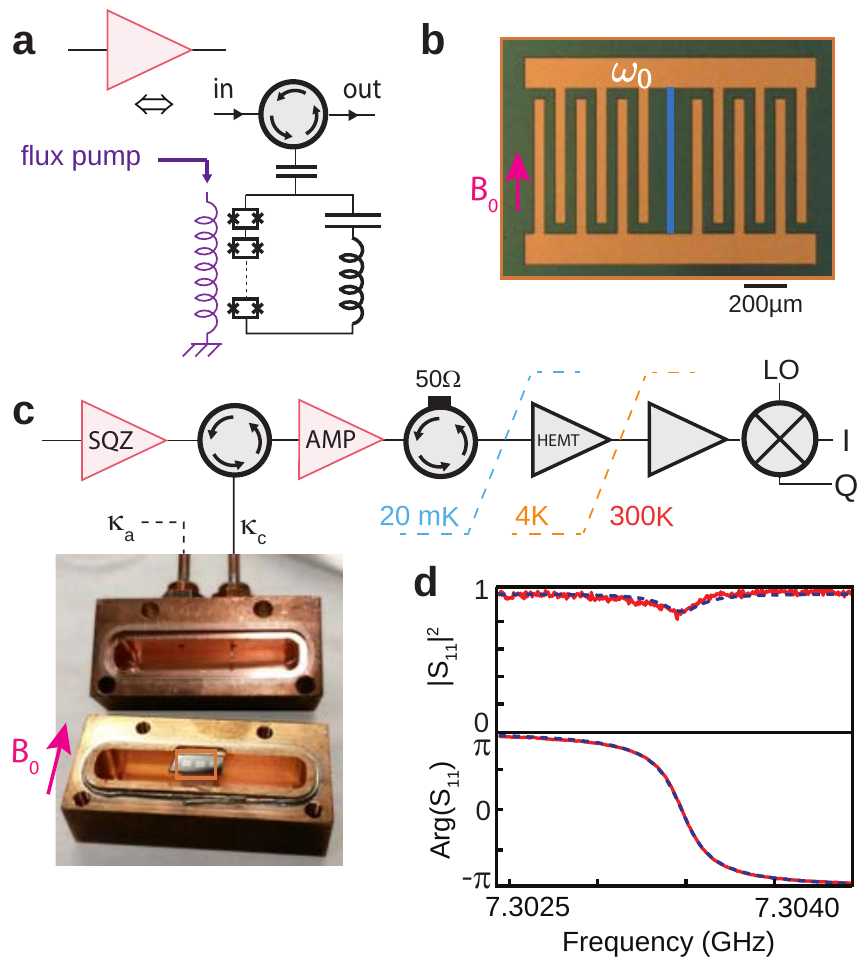}
\caption{Experimental setup. (\textbf{a}) The JPAs providing the squeezed  vacuum and the noiseless amplification are superconducting LC resonators containing a SQUID array, tuned to a frequency close to the ESR cavity frequency $\omega_0$ by the application of a d.c. flux bias to the SQUID loops. Modulating this flux at twice the resonator frequency by application of a pump tone yields parametric gain at $\omega \simeq \omega_0$ for the signals reflected off the JPA. (\textbf{b}) The ESR cavity, whose optical micrograph is shown, is an aluminum lumped-element LC resonator of frequency $\omega_0$ patterned on top of a silicon sample containing the spins. A magnetic field $B_0$ is applied parallel to the sample surface and to the resonator inductor (in blue) to tune the spin frequency. Only spins in the immediate vicinity of the inductor are detected. (\textbf{c}) The resonator is enclosed in a copper box holder and is capacitively coupled to the measurement line with a constant $\kappa_C$ via an antenna fed through the 3D copper sample holder, thermally anchored at 20\,mK. A second port, much less coupled (constant $\kappa_A \ll \kappa_C$), is used for characterization (see text). Squeezed microwaves at $\omega_0$ are generated by a first JPA denoted SQZ, routed onto the resonator via a circulator, and the reflected signal is noiselessly amplified by a second JPA denoted AMP. Both are operated in the degenerate mode, and pumped at $2\omega_0$ with respective phases $\phi_S$ and $\phi_A$. Further amplification is provided at \ay{$4$\,K} by a High Electron Mobility Transistor amplifier and at \ay{ $300$\,K}. Homodyne demodulation at the signal frequency yields the quadratures $I(t)$ and $Q(t)$. (\textbf{d})  Measured reflection coefficient $|S_{11}^2|$ (red line, blue line is a fit) yielding $\omega_0/2\pi=7.3035$\,GHz, $\kappa_C=1.6 \times 10^{6}\, \rm{s}^{-1}$ and $\kappa_A + \kappa_L=6 \times 10^{4}\,\rm{s}^{-1}$, $\kappa_L$ being the resonator internal loss rate. We determine $\kappa_A = 3 \times 10^{3}\, \rm{s}^{-1}$ by measuring the full resonator scattering matrix (not shown).  \label{fig2}}
\end{figure}

\pb{We now turn to the experimental implementation of this proposal, starting with the description and the characterization of the squeezed microwave source.} Squeezing and noiseless amplification are achieved by the same type of device : a flux-pumped JPA operated in the degenerate mode, denoted SQZ for the squeezer and AMP for the amplifier. The JPA consists of a single-port resonator of frequency close to $\omega_0$ containing a SQUID array ~(see Fig.~\ref{fig2}a). The magnetic flux threading each SQUID loop is modulated by a pump tone at frequency $2\omega_0$ with a phase $\phi$ leading to a resonator frequency modulation $\propto \cos [ 2( \omega_0 t + \phi)]$~\cite{Zhou.PhysRevB.89.214517(2014)}. Parametric amplification with amplitude gain $G$ occurs for input signals $V \cos \omega_0 t$ if $ \phi = - \pi / 4$, and de-amplification with gain $1/G$ if $ \phi = + \pi / 4$. The SQZ is a JPA acting on the vacuum at its input, generating a squeezed vacuum state with a variance reaching $1/(4G^2)$ on its squeezed quadrature and an average photon number $N = (G^2 + G^{-2}-2)/4$ (see Supplemental Material).

As explained in Section~\ref{sec:theory}, squeezing is very sensitive to microwave losses. It is thus important to characterize the squeezed state with a setup that contains all elements used for the magnetic resonance experiment (described in Section IV), including the ESR cavity. This ESR cavity consists of a high-quality-factor superconducting LC resonator patterned on top of a silicon sample. It is enclosed in a copper box holder and connected to the measurement line by capacitive coupling to an antenna (see Fig.~\ref{fig2}c), whose length sets the coupling constant $\kappa_C$. The squeezed vacuum generated by SQZ is sent into the ESR cavity input via a circulator which routes the reflected field into AMP. Further amplification stages include a semiconductor high-electron-mobility transistor (HEMT) amplifier at 4\,K as well as room-temperature amplifiers. Transmission and reflection coefficients can be measured with a network analyzer. For phase-sensitive measurements, a microwave signal at the cavity frequency $\omega_0$ is sent into the JPA. After amplification, the output signal is demodulated by mixing with a local oscillator also at $\omega_0$, yielding time traces of the quadratures $I(t)$ and $Q(t)$ that are digitized with a $300$\,kHz bandwidth. More details on the setup can be found in the Supplemental Material and in~\cite{bienfait2015reaching}.

\begin{figure}[!htbp]
\includegraphics[width=88mm]{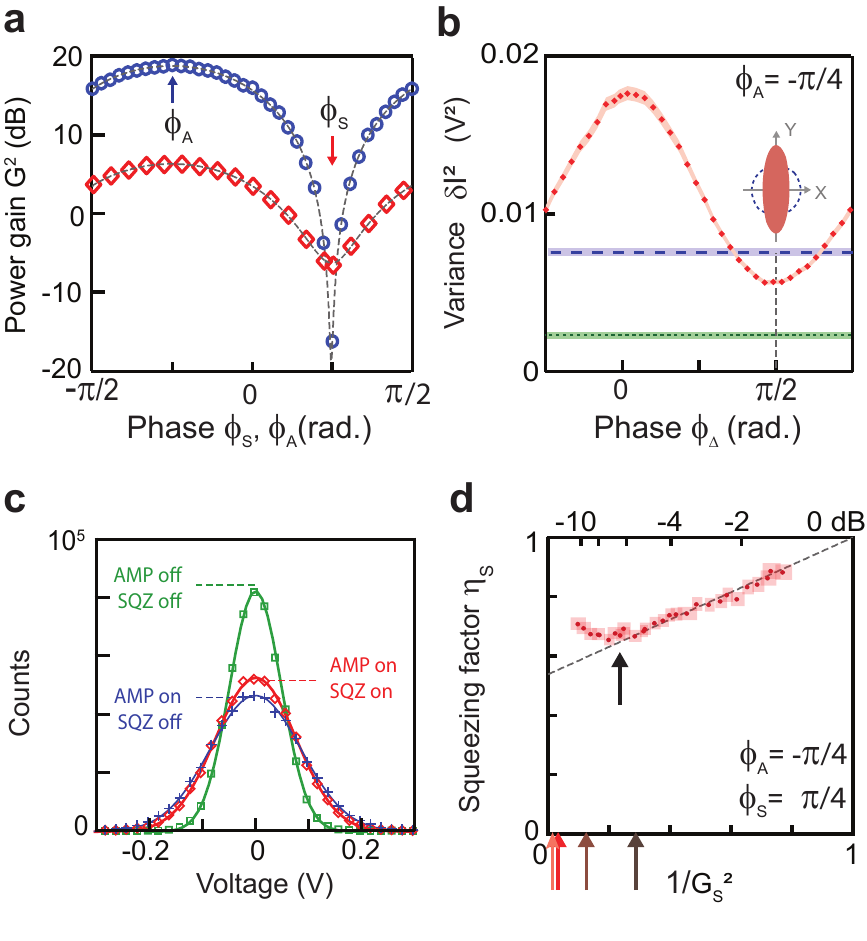}
\caption{Characterisation of the prepared squeezed vacuum state. (\textbf{a}) Gain of SQZ and AMP as a function of their pump phase ($\phi_{S,A}$, respectively) for the chosen pump amplitude settings, leading to a power gain $G_S^2 = 6$\,dB and $G_A^2 = 20$\,dB \ay{for $\phi_{S,A}= - \pi/4$}. Optimal values for the SQZ and AMP pump phases are indicated with arrows. (\textbf{b}) The variance $\delta I^2$ in the noise is plotted as a function of the difference between the SQZ and AMP pump phases $\phi_\Delta = \phi_S - \phi_A$, $\phi_A$ being set at its optimal value. Data with AMP and SQZ both on (red open squares) are compared to those obtained with AMP on and SQZ off (blue dashed line), and with AMP and SQZ both off (green dashed line). \ay{Shaded areas represent the 5$\sigma$ measurement uncertainty}. Squeezing is obtained for the optimal setting $\phi_{\Delta} = \pi / 2$. (\textbf{c}) Noise histograms obtained using the optimal phases obtained above, for AMP and SQZ both off (green open squares) in which case the noise is determined by the HEMT amplifier, AMP on and SQZ off (blue crosses) in which case the noise is the sum of the HEMT and the amplified vacuum fluctuations, and AMP and SQZ both on (red open symbols) in which case the fluctuations are reduced below the vacuum level. Gaussian fits for each are also shown (curves). (\textbf{d}) Using the optimal phase settings, the squeezing factor (see main text) $\eta_S$ is measured as a function of the SQZ power gain (red dots, \ay{rectangles represent the 5$\sigma$ measurement uncertainty}) \ay{by varying the SQZ pump power}. A linear fit (dashed line) for the low-gain part of the curve indicates the microwave losses between SQZ and AMP to be $\eta_{\rm loss} = 0.54$. The black arrow indicates the SQZ gain selected in the experiment. \label{fig3}}
\end{figure}

To characterize the squeezed microwave state, we keep the spins detuned from the ESR resonator by working at $B_0=0$\,mT. We show in Fig.~\ref{fig3}a the effect of pump phase ($\phi_S$ and $\phi_A$) on the power gain (\ay{$G_S^2$}  and \ay{$G_A^2$}) for SQZ and AMP, respectively. The gains vary sinusoidally as expected, with $G_S^2 = 6$\,dB and $G_A^2=18$\,dB for the chosen pump amplitude settings. In the remainder of this work, the local oscillator phase is set such that the quadrature maximally amplified by AMP is $I(t)$. Note that the detection bandwidth is much smaller than the $3$\,MHz bandwidth of both JPAs. The variance $\delta I^2$ is shown in Fig.~\ref{fig3}b as a function of the relative phase between the SQZ and AMP pump signals $\phi_{\Delta} = \phi_S - \phi_A$, with no signal at the input. As demonstrated in~\cite{mallet_quantum_2011}, $\delta I^2$ depends on $\phi_{\Delta}$, allowing us to experimentally determine the optimal squeezing condition $\phi_{\Delta} = \pi / 2$. Statistical distributions of $I(t)$ are shown in the form of histograms in Fig.~\ref{fig3}c for this optimal condition. We find that the variance in the total output noise ($\delta I^2_{\mathrm{on}}$) is reduced $1.2$\,dB below that observed with the SQZ pump off ($\delta I^2_{\mathrm{off}}$), $\delta I^2_{\mathrm{on}} = 0.75 \delta I^2 _{\mathrm{off}} $ (see Figs.~\ref{fig3}b and c). 

In order to determine whether this reduced noise is indeed below the vacuum fluctuations level, it is necessary to determine how close $\delta I^2_{\mathrm{off}}$ is to the vacuum fluctuations value $\delta I^2_0$. It is indeed well-known that the temperature of the cavity field may differ from the sample temperature of $10$\,mK, due to leakage of thermal radiation from higher temperature stages. Calibration measurements (reported in the Supplemental Material) were performed using a transmon qubit and enable us to put an upper bound of $\bar{n}=0.1$ thermal photon present in the mode. Since in a thermal state the variance of a quadrature $\hat{X}$ is given by $\langle \delta X^2 \rangle = (1 + 2 \bar{n}) / 4$, we can state that in our experiment $\delta I^2_0<\delta I^2_{\mathrm{off}}<1.2 \delta I^2_0$. We thus come to the conclusion that the measured noise with SQZ on \ay{is lower than the vacuum fluctuations by at least $10\%$ and at best by $25\%$}, proving that the produced squeezed state is truly in the quantum regime.

This figure is limited by several factors, one of them being the sensitivity of squeezed states to losses, as discussed in Section\,~\ref{sec:theory}. For our experiment, the relevant microwave losses are those between SQZ and AMP, which include the insertion loss of circulators and cables, and internal losses of the SQZ and AMP devices as well as of the ESR resonator. Care was taken to minimize these losses; in particular, the coupling rate of the resonator to the output waveguide, $\kappa_C = 1.2 \cdot 10^6\,\mathrm{s}^{-1}$, was purposely set to be much larger than the internal loss rate of resonator, $\kappa_L =  6 \cdot 10^4\,\mathrm{s}^{-1}$ so that the losses in reflection are below $1$\,dB \pb{as shown in Fig.~\ref{fig2}d}.

To quantify these losses, we measure the squeezing factor $\eta_S\equiv(\delta I^2 _{\mathrm{on}} -  \delta I^2 _{\mathrm{bg}})/( \delta I^2_{\mathrm{off}} -  \delta I^2 _{\mathrm{bg}})$ as a function of $G_S^2$ (see Fig.~\ref{fig3}d), $\delta I^2 _{\mathrm{bg}}$ being the variance in the background noise obtained when both SQZ and AMP are switched off. With this definition, $\eta_S$ measures only the quantum noise reduction due to squeezing; according to Eq.~\ref{eq:etaloss} one expects $\eta_S = \eta_{\mathrm{loss}} + (1-\eta_{\mathrm{loss}})/G_S^2$. At low gain ($\lesssim5$\,dB), $\eta_S$ is indeed measured to depend linearly on $G_S^{-2}$. A linear fit yields \ay{$\eta_{\rm loss} = 0.54 $.}

\begin{figure}[!htbp]
\includegraphics[width=88mm]{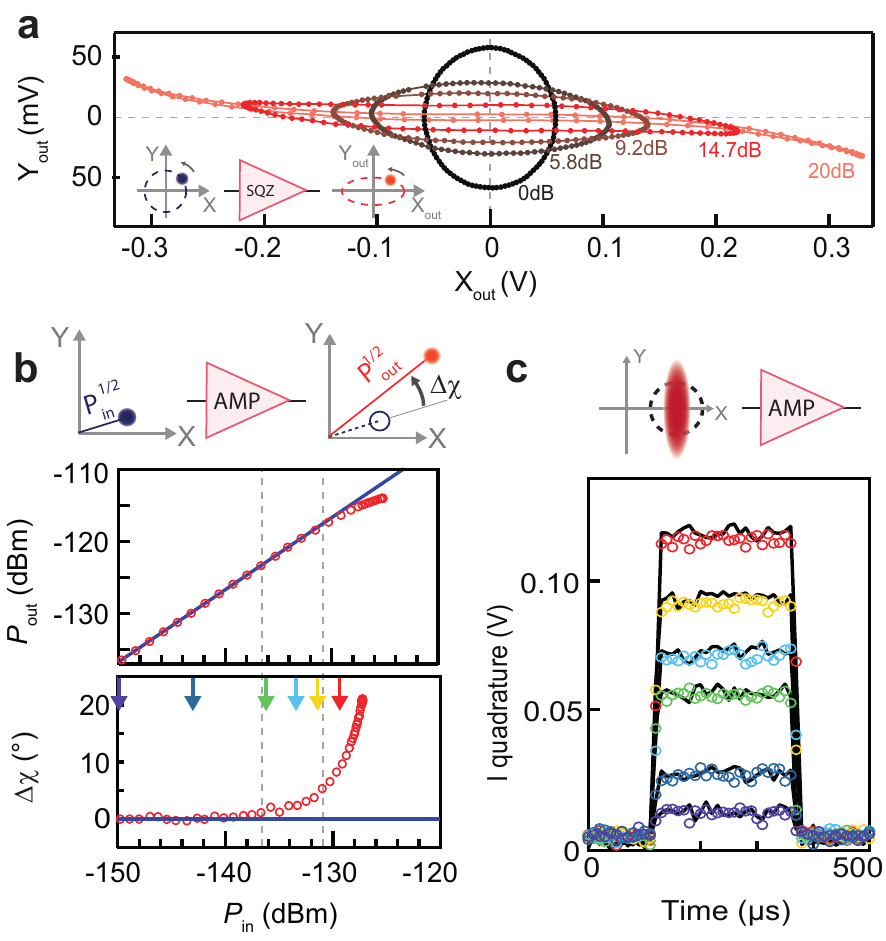}
\caption{Limitations induced by JPA saturation. (\textbf{a}) Output quadratures $X_{\rm{out}}$ and $Y_{\rm{out}}$ measured for weak coherent signals sent to the SQZ with input phases $\phi_S$ spanning the whole interval between $0$ and $2\pi$ for 4 different gains $G_S^{-2}$ indicated by the arrows in Fig.\,\ref{fig3}d) \ay{and set using different pump powers (black curve corresponds to SQZ pump off)}. Note that these data were obtained in a separate calibration run in which the ESR cavity was removed~\cite{BienfaitPhD}. (\textbf{b}) Measured output power and phase of a signal at $\omega_0$ as a function of its input power $P_{\mathrm{in}}$ after amplification by the AMP device, in the same conditions as in panel (a) but with the JPA operated in the non-degenerate mode by detuning the pump by $300$\,kHz from $2\omega_0$. The output power depends linearly on $P_{\mathrm{in}}$ as long as $P_{\mathrm{in}}<-131$\,dBm (blue lines is a linear fit), while the phase shift is zero (blue line) only for $P_{\mathrm{in}}<-137$\,dBm. (\textbf{c}) Time trace of the $I$ quadrature of a weak microwave pulse at $\omega_0$ sent via the $\kappa_A$ port of the ESR cavity, measured with SQZ off (open circles) and SQZ on (black traces) for different input powers indicated by the arrows in panel b. The traces have been averaged $4000$ times. Above $-136$\,dBm, deviations appear between the SQZ on and SQZ off curves due to AMP saturation. \label{fig4}}
\end{figure}

At higher gain, a departure from linearity is observed, with an increase of the variance. To investigate this phenomenon, we measure the SQZ response to a coherent input signal having an amplitude corresponding to the root-mean-square vacuum fluctuations, with the SQZ operated in degenerate mode. Varying this signal phase $\phi$ from $0$ to $2\pi$, the output quadratures $(X_{\rm{out}}(\phi),Y_{\rm{out}}(\phi))$ mimic the shape of the produced squeezed vacuum, as seen in Fig.~\ref{fig4}a. At small or moderate gains ($G_S^2< 6$\,dB), an ellipse is observed with its small-axis projection scaling as $G_S^{-2}$. For gains larger than $10$\,dB however, the ellipse becomes strongly distorted, which explains the increase of the squeezed quadrature variance at high gain observed in Fig.~\ref{fig3}d. We attribute this ellipse distorsion phenomenon to cubic or quartic non-linearities in the parametric amplifier Hamiltonian, arising from higher-order terms in the expansion of the Josephson junction potential energy~\cite{wustmann_parametric_2013,boutin2017effect}. We therefore choose to set the SQZ pump power such that $G_S^2=6$~dB, which yields the largest amount of squeezing as seen in Fig.~\ref{fig3}d.

The non-linearity of the JPA devices also affects the operation of the noiseless amplifier AMP, by causing power-dependent phase shifts and saturation of the output power, as seen in Fig.~\ref{fig4}b. Power-dependent phase shifts are particularly detrimental for our experiment: squeezed states have a higher power than the vacuum, which implies that the echo signal may be phase-shifted when the squeezing is turned on, leading to a reduced output amplitude. We illustrate this effect by sending a small coherent pulse onto the ESR resonator via an additional port (see Fig~\ref{fig2}c) whose coupling rate to the ESR resonator $\kappa_A \ll \kappa_L,\kappa_C$ is negligible. The phase of the coherent pulse and the SQZ pump are set so that the pulse is detected on the $I$ quadrature and is aligned with the squeezed vacuum. Fig~\ref{fig4}c shows the recorded time traces $I(t)$ with squeezing switched on and off for different input powers. While for powers below $-136$\,dBm the amplitudes observed with squeezing on and off are identical as desired, at larger input powers there is a difference by a few percents. To avoid this effect in the experiment described in the next section, we limit the power of the spin-echo signal well below this threshold.

\section{ESR spectroscopy in the presence of squeezing}
\label{sec:ESRWithSqueezing}

\begin{figure}[h!]
\centering
\includegraphics[width=100mm]{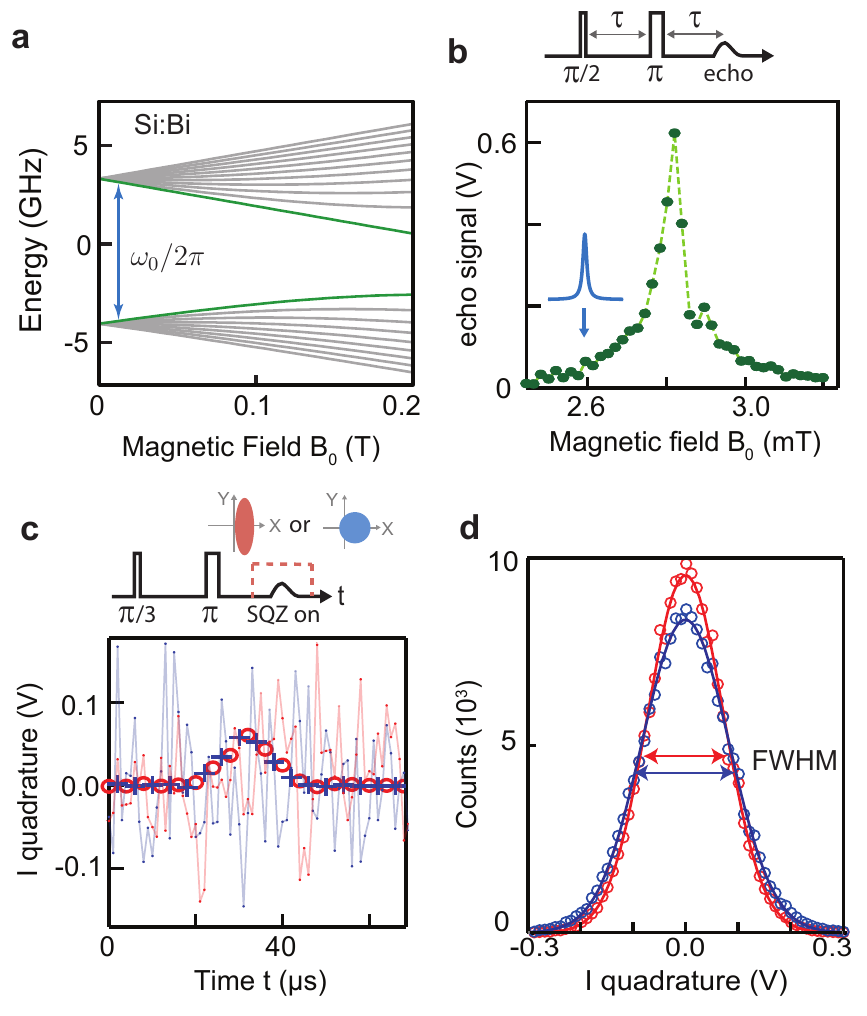}
\caption{Squeezing-enhanced spin-echo detection. (\textbf{a}) Energy of the $20$ levels of bismuth donors in silicon as a function of $B_0$ (grey lines). The transition between the levels indicated in green is used in the experiment. (\textbf{b}) Hahn-echo detected magnetic field sweep, showing the bismuth donor resonance line. Blue arrow indicates the field chosen in the rest of the experiment, blue Lorentzian curve indicates the fraction of spins that are within the cavity resonance. (\textbf{c}) Echo signals observed with SQZ off (blue) and on (red) for a single shot (lines) and averaged over $2500$ traces (symbols) confirm the signal intensity is identical. SQZ was switched on only during a short $\Delta t = 200\,\mu \mathrm{s}$ window around the echo emission time (dashed rectangle in the pulse sequence). The excitation pulse angle is chosen to be $\approx \pi / 3$ in order to avoid saturation effects (see main text). (\textbf{d}) Histograms of the noise around the average signals of panel \ay{c} measured with 2500 single-shot traces acquired on a $70 \, \mu \mathrm{s}$-time-window centered on the echo, with SQZ off (blue) and on (red) (see Supp. Mat.), and corresponding Gaussian fits (dashed curves). Standard deviations are $0.0858 \pm 2 \cdot 10^{-4}$\,V for SQZ off and $0.0748 \pm 2 \cdot 10^{-4}$\,V for SQZ on, confirming a reduction in the noise accompanying the spin-echo signal when SQZ is on. \label{fig5}}
\end{figure}

The spins used in the experiment are provided by bismuth ($^{209}$Bi) donors implanted in the silicon sample, which has been isotopically enriched in the nuclear-spin-free $^{28}\mathrm{Si}$. At low magnetic fields, the strong hyperfine interaction between the $S=1/2$ electron and the $I=9/2$ nuclear spins yields multiple allowed ESR-like transitions around $7.37$\,GHz (see Fig.~\ref{fig5}a); we work here on the lowest frequency transition. More details on the characterization of this sample can be found in~\cite{Wolfowicz.NatureNano.8.561(2013),Weis.APL.100.172104(2012)}.

Figure~\ref{fig5}b shows this spin resonance line, obtained by measuring the spin-echo intensity as a function of the magnetic field $B_0$ (with SQZ off). The expected resonance is found around $B_0 = 2.8$\,mT (see Figs.~\ref{fig5}a and b), with the $0.1$\,mT linewidth primarily due to strain exerted by the aluminium wire on the underlying silicon substrate~\cite{Pla.Arxiv.1608.07346}. \ay{The spin linewidth is considerably broader ($\times 30$) than the resonator bandwidth. As a consequence, only a narrow subset of spins is excited at each magnetic field $B_0$ and contributes to the echo signal. For our $5$-$\upmu$s $\pi/2$ excitation pulse, we estimate the excitation bandwith to $100$\,kHz, justifying our choice of a $300$-kHz digitization bandwitdh.} Rabi oscillations (obtained by sweeping the power of the \ay{$10\,\upmu$s rectangular} refocusing pulse in the Hahn echo sequence) were used to calibrate the pulses for subsequent experiments~\cite{bienfait2016Purcell}. To avoid saturation of AMP as explained in the previous section, we purposely set the field far away from the maximum of the spin resonance line at $B_0=2.6$\,mT (see Fig.~\ref{fig5}b), and we use an echo sequence of the form $\theta-\tau-\pi-\tau-{\rm echo}$ in which the usual $\pi/2$ Rabi angle of the Hahn echo is replaced by a lower Rabi angle $\theta$ (\ay{the $\theta \simeq \pi/3$ rotation is realized via a $3$-$\upmu$s-rectangular pulse}) with $\tau=200~\mu$s.

An echo was then recorded in the two following conditions : SQZ off; and SQZ pump switched on for a time window of $200\, \mu \mathrm{s}$ centred around the echo emission time (pulsing the squeezed state generation was found to be crucial for the success of the experiment, for reasons explained in the next section). The phases of the excitation and refocusing pulses were set in such a way that the echo signal was produced entirely on the $I$ quadrature aligned with the squeezed vacuum. Time traces of the digitized $I(t)$ quadrature are shown in Fig.~\ref{fig5}c, with the echo barely visible in single-shot traces. After averaging, the spin-echo amplitude appears to be identical for SQZ on and off, confirming that the saturation effects mentioned in Section III were avoided. \ay{The $20$-$\upmu$s echo duration arises from the excitation bandwidth}. Histograms of the noise during the echo emission (Fig.~\ref{fig5}d) show that the data obtained with SQZ on exhibit less noise than with SQZ off --- indeed, the similarity between these distributions and those obtained with no spin-echo signal (Fig.~\ref{fig3}c) confirm that quantum fluctuations are the primary noise source in the spin echo measurements. In both cases the variance is reduced by a factor of $0.75$ when SQZ is on, in agreement with the theoretical analysis presented in Section\,~\ref{sec:theory}. As the noise reduction is obtained while maintaining constant spin-echo signal amplitude, this demonstrates that the sensitivity of magnetic resonance detection is enhanced using quantum squeezing.

\section{Discussion and conclusion}
\label{sec:DiscussionConclusion}


\subsection{Applicability of the scheme}

\pb{We now discuss to what extent our proof-of-principle demonstration of squeezing-enhanced magnetic resonance detection can be improved to be of practical use. The measured noise reduction of $1.2$\,dB provided by the squeezed state injection is limited by the finite value of the input field squeezing factor $\eta_S$, as seen in Fig.~\ref{fig3}d. As explained in Sec.~\ref{sec:squeezing}, this is due in our experiment to two distinct phenomena. The first one is the non-linearity of both the SQZ and the AMP parametric amplifiers, which puts a lower bound on $\eta_S$ and limits the maximum spin signal that can be amplified and detected. All these issues can be solved by using other JPAs with up to $30$\,dB higher saturation power than our design~\cite{Mutus.APL.104.26(2014),roy2015broadband}. Keeping everything else unchanged in our experiment, we estimate that using these optimized amplifiers would have resulted in a noise reduction of $\approx 3$\,dB. The second factor limiting the squeezing-induced noise reduction is the presence of microwave losses on the path over which the squeezed states propagates; those are due to cables, circulators, and resonator internal losses. As explained in Sec.~\ref{sec:squeezing}, they add up to $\approx 3$\,dB in our setup, but simple improvements (minimizing cable length, using a lower number of circulators, and a resonator of lower internal losses) could bring this figure down to $1$\,dB. These straightforward improvements realistically lead to a noise reduction on the squeezed state quadrature by a factor $5$ below the vacuum level. The resulting reduction of the measurement time by a factor $5$ with unchanged signal-to-noise ratio is clearly relevant for practical applications. To reduce the losses below $1$\,dB, more radical setup changes would probably be needed, such as integrating the squeezer, the circulator, the ESR resonator and the amplifier on a single chip. While such a complex quantum integrated circuit has never been achieved so far, promising steps in that direction have already been 
taken, with in particular several recent demonstrations of on-chip superconducting circulators~\cite{Kerckhoff.PhysRevApplied.4.034002,Sliwa.PhysRevX.5.041020,chapman2017widely,Lecoq.PhysRevApplied.7.024028(2017)}.}

\pb{When considering the practical relevance of our scheme and results for magnetic resonance spectroscopy, one point deserves attention. Because microwave losses must be minimized to preserve the degree of squeezing of the field, the resonator should be largely over-coupled to the measurement line (i.e. $\kappa_C \gg \kappa_L$). However the measurement sensitivity of a spin-echo detection scales like $\kappa^{-1/2}$~\cite{bienfait2015reaching}, $\kappa = \kappa_C + \kappa_L$ being the total resonator loss rate. Increasing $\kappa_C$ thus reduces the measurement sensitivity in absence of squeezing. In our experiment for instance, we estimate that $N_{min} = 1.3\times 10^4$ spins can be detected with a signal-to-noise ratio of $1$ per echo without squeezing (see online Supplemental Material), a factor $7$ larger than what was achieved with a critically-coupled resonator~\cite{bienfait2015reaching}. One can thus wonder whether it is really more desirable for ESR measurements to use an over-coupled resonator with squeezed microwaves sent onto its input, instead of a critically-coupled resonator without any squeezing. Focusing exclusively on spin sensitivity however neglects the fact that most electron spin species in general have a linewidth which is much broader than the $\approx 10-50$\,kHz linewidth of a critically-coupled superconducting resonator~\cite{bienfait2015reaching}. In this case, the lower intrinsic spin sensitivity of an over-coupled resonator is compensated by the larger number of spins measured, resulting in unchanged signal-to-noise ratio for spin-echo measurements. In other words, it is desirable to match the resonator linewidth to the spin linewidth; with superconducting resonators this amounts to over-coupling the resonator to the measurement line, exactly as required for the squeezing enhancement. Note that lowering the resonator quality factor comes with other advantages, such as the possibility of applying large-bandwidth control pulses. To sum up, squeezing-enhanced ESR spectroscopy is well suited to measure spins whose linewidth is comparable or larger than the one of an over-coupled superconducting resonator ($\approx 1$\,MHz, i.e. $0.03$\,mT), which is the case in our experiment as well as for many spin species~\cite{SchweigerEPR(2001)}.}

\begin{figure}[h!]
\centering
\includegraphics[width=90mm]{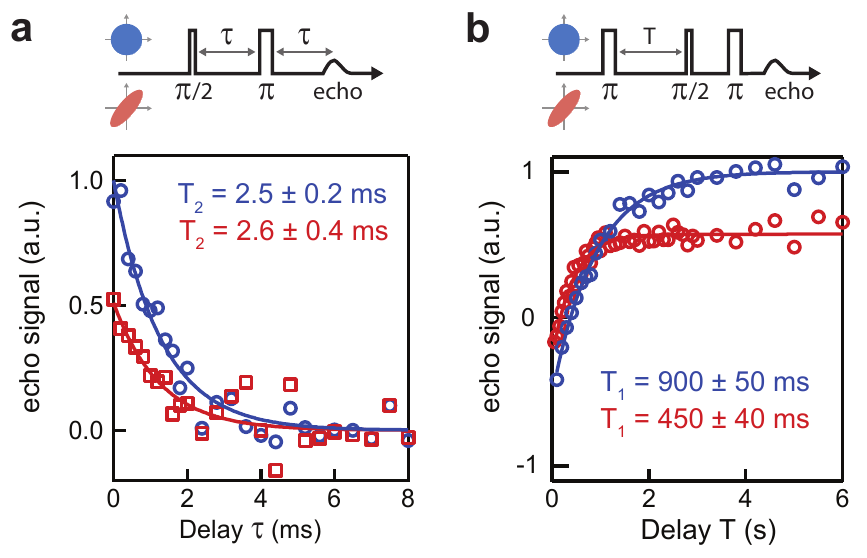}
\caption{Influence of squeezing on the spin coherence times. (\textbf{a}) Coherence time $T_2$ measured with a Hahn echo sequence for SQZ off (blue circles) and on (red squares). Contrary to the experiments in Fig.~5, SQZ is now switched on or off for the entire experimental sequence. The integrated echo signal is plotted as a function of the delay $\tau$ between the $\pi/2$ and $\pi$ pulses. Exponential fits (solid lines) yield $T_{2,off} = 2.5 \pm 0.2\, $ms and $T_{2,on} = 2.6 \pm 0.4\, $ms. (\textbf{b}) Energy relaxation time $T_1$ with SQZ on (blue circles) and off (red circles). Exponential fits (solid lines) yield $T_1 = 900 \pm 50$\,ms with SQZ off and $T_1 = 450 \pm 40$\,ms with SQZ off. Both $T_1$ and $T_2$ curves (panels a and b) have their amplitude reduced by $\approx 0.5$ with SQZ on, indicating reduced spin polarisation in the steady-state when the SQZ is continuously switched on.
\label{fig6}}
\end{figure}

\subsection{Ultimate limits to the sensitivity and squeezing back-action on the spin dynamics}

\pb{After discussing the applicability of our scheme, we now wish to address a more fundamental question. Supposing that an ideal squeezed state of arbitrary squeezing factor could be sent onto the cavity so that the purely electromagnetic contribution to the total noise would be completely suppressed. It is then worthwhile to investigate which other physical mechanisms would ultimately limit the sensitivity. Those can be deduced from Eq.~\ref{eq:noise}, which shows that in the limit where $\delta X_{\mathrm{in}} \rightarrow 0$, a finite variance is maintained, $\delta X^2 = |l(0)|^2 \delta X_{\mathrm{bath}}^2 + |t(0)|^2 \delta X_{\mathrm{spin}}^2$. In an ideal experiment, the cavity internal losses may be suppressed so that the first term is negligible; however the second term describes noise emitted by the spins (as observed experimentally in~\cite{Sleator.PhysRevLett.55.1742(1985)}), and is thus unavoidable. }Negligible in our experiment, this contribution becomes relevant in the limit where the squeezing factor $\eta_S$ becomes comparable to the ensemble cooperativity $4 C(0)$ as seen in Section \ref{sec:theory}. Additional measurements (reported in the Supplemental Material) yield $C(0) = 0.002$, which implies that spin noise would be a limitation for $20$\,dB squeezing, corresponding to a maximum gain in sensitivity by a factor $\simeq 10$.

Another fundamental effect disregarded so far concerns the effect of squeezed radiation on the spin dynamics, which may in certain cases lead to a reduction of the echo signal. Indeed, in steady-state, squeezed radiation incident on a two-level system modifies its relaxation and coherence times as well as its average polarisation, as predicted in Ref~\cite{Gardiner.PhysRevLett.56.1917(1985)} and observed in recent experiments with superconducting qubits~\cite{Murch.Nature.502.211(2013),toyli2016resonance}. Note that squeezing does not affect the damping rates of a harmonic oscillator, which explains its absence in the analysis of Section\,~\ref{sec:theory} where the spins are modelled as a collection of oscillators~\cite{kiilerich2017relaxation}.

To investigate experimentally these effects, we measure spin coherence and relaxation times with SQZ turned off or on during the entire experimental sequence. $T_2$ is found to be unaffected by squeezing (see Fig.~\ref{fig6}a), because decoherence occurs by non-radiative processes such as dipolar interactions~\cite{tyryshkin_electron_2012}. Energy relaxation on the other hand has been shown to be caused by spontaneous emission of microwave photons through the cavity (the Purcell effect) with a rate $T_1^{-1} = 4 g^2 / \kappa$~\cite{bienfait2016Purcell}, $g$ being the coupling of a single spin to the radiation field as defined in Section\,~\ref{sec:theory}. Being of radiative origin, the effective $T_1$ should be modified by the squeezed microwave injection. Accordingly, it is found to decrease from $0.9$\,s to $0.45$\,s when squeezing is continuously switched on (see Fig.~\ref{fig6}b), with an overall echo amplitude diminished by the same factor $2$, revealing the expected reduced spin polarisation. The reduction factor on both $T_1$ and polarisation is predicted by Gardiner to be $1+2N$~\cite{Gardiner.PhysRevLett.56.1917(1985)}, yielding $N\approx 0.5$ in our experiment, compatible with the chosen SQZ gain $G_S^2 = 6$\,dB as well as with the squeezed state characterisation by homodyne detection shown in Fig.~\ref{fig3}.

Squeezing-induced spin depolarisation was avoided in the data shown in Fig.~\ref{fig3}c because SQZ was only switched on for a short time window $\Delta t = 200 \, \mu \mathrm{s}$ around the echo, much smaller than the depolarisation time which is of order $\approx T_1 / (1+2N)=0.45$\,s in our experiment. This strategy can only be applied if the depolarisation time is longer than the echo duration $T_{\mathrm{E}}$, i.e. if the squeezing parameter $\eta_S \approx 1/N > 8 g^2 T_{\mathrm{E}} / \kappa$. It is interesting to note that  $8 g^2 T_{\mathrm{E}} / \kappa \approx 4 C(0) / N_{\mathrm{spins}}$, which is the \textit{single-spin} cooperativity and is therefore much smaller than the ensemble cooperativity $4C(0)$ as long as the ensemble contains a large number of spins $N_{\mathrm{spins}} \gg 1$. Spin noise is therefore expected to limit the achievable sensitivity gain much earlier than spin depolarization, provided the squeezed state generation is pulsed as in the present experiment. For our experimental parameters, $T_1 / T_{\mathrm{E}} \approx 10^{-5}$, so that squeezing-induced spin depolarization would not be an issue unless $50$\,dB squeezing is reached, instead of the $20$\,dB limit found for spin noise.

\subsection{Conclusion}

In conclusion, we have presented a proof-of-principle demonstration of squeezing-enhanced magnetic resonance detection. While the fundamental limitations to this scheme deserve further study, \pb{our results could likely be improved using present-day technologies to gain up to a factor $5$ in measurement time, reaching the point at which the method becomes practically relevant for magnetic resonance measurements.} Besides improving the degree of squeezing, future work could investigate the use of other non-classical states such as Schr\"odinger-cat states in magnetic resonance, which might bring even larger sensitivity gains~\cite{giovannetti2004quantum,facon2016sensitive}.

\begin{acknowledgments}
We acknowledge technical support from P. S{\'e}nat, D. Duet, J.-C. Tack, P. Pari, P. Forget, as well as useful discussions within the Quantronics group. We acknowledge support of the European Research Council under the European Community's Seventh Framework Programme (FP7/2007-2013) through grant agreements No. 615767 (CIRQUSS), 279781 (ASCENT), and 630070 (quRAM), and of the C'Nano IdF project QUANTROCRYO. J.J.L.M. is supported by the Royal Society. T.S. was supported by the U. S. Department of Energy under contract DE-AC02-05CH11231. A.H.K. and K.M. acknowledge support from the Villum Foundation.
\end{acknowledgments}


\newpage
\widetext
\begin{center}
\textbf{\large Supplementary Material: Magnetic Resonance with Squeezed Microwaves}
\end{center}
\setcounter{equation}{0}
\setcounter{figure}{0}
\setcounter{table}{0}
\setcounter{page}{1}
\makeatletter
\renewcommand{\theequation}{S\arabic{equation}}
\renewcommand{\thefigure}{S\arabic{figure}}

%

\section{Theory details}
\subsection{Squeezing and amplification by a Josephson parametric amplifier}
In the experiment, we use Josephson parametric amplifiers (JPA) to both produce a squeezed vacuum input field incident on the resonator and to amplify a single quadrature of the emitted radiation in a noiseless manner. For a general review on parametric amplification, see~\cite{Gardiner.QuantumNoise}. Here we establish the formalism, focusing on the limit of broad band squeezing. 

The transformation of the field quadrature operators by a JPA employed in the degenerate mode is described by an amplitude gain factor $G$, 
\begin{align}
\begin{split}
\label{eq:ampl-gain}
\hat{X}_{\mathrm{out}}&=G \hat{X}_{\mathrm{in}} \\
\hat{Y}_{\mathrm{out}}&=G^{-1} \hat{Y}_{\mathrm{in}}.
\end{split}
\end{align}
The amplification, that we apply to the output from the ESR resonator, is hence unitary and maintains the signal-to-noise ratio~\cite{PhysRevD.23.1693(1981)}.
Equations\,(\ref{eq:ampl-gain}) are equivalent to a transformation of the field annihilation and creation operators ($\hat{X}=\frac{1}{2}(\hat{a}+\hat{a}^\dagger),\ \hat{Y}=\frac{1}{2i}(\hat{a}-\hat{a}^\dagger)$),
\begin{align}
\begin{split}\label{eq:field-bog}
\hat{a}_{\mathrm{out}}&=\frac{G+G^{-1}}{2}  \hat{a}_{\mathrm{in}} + \frac{G-G^{-1}}{2}\hat{a}_{\mathrm{in}}^{\dagger}\\
\hat{a}_{\mathrm{out}}^{\dagger}&=(\hat{a}_{\mathrm{out}})^{\dagger}.
\end{split}
\end{align}
Assuming the broadband limit and applying the JPA to a vacuum or a thermal state, Eqs.~\ref{eq:field-bog} lead to the temporal correlations of the squeezed output field,
\begin{align}
\begin{split}
 \label{eq:field-temp}
\langle \hat{a}^{\dagger}_{\mathrm{out}}(t) \hat{a}_{\mathrm{out}}(t')\rangle &=\left(\frac{G+G^{-1}}{2}\right)^2 \langle \hat{a}^{\dagger}_{\mathrm{in}}(t) \hat{a}_{\mathrm{in}}(t') \rangle+ \left(\frac{G-G^{-1}}{2}\right)^2 \langle \hat{a}_{\mathrm{in}}(t) \hat{a}^{\dagger}_{\mathrm{in}}(t')\rangle  \equiv N \delta(t-t') \\
\langle \hat{a}_{\mathrm{out}}(t) \hat{a}_{\mathrm{out}}(t')\rangle &=\frac{G+G^{-1}}{2}\frac{G-G^{-1}}{2} \left(\langle \hat{a}^{\dagger}_{\mathrm{in}}(t) \hat{a}_{\mathrm{in}}(t') \rangle+\langle \hat{a}_{\mathrm{in}}(t) \hat{a}^{\dagger}_{\mathrm{in}}(t')\rangle\right) \equiv M\delta(t-t').
\end{split}
\end{align}
The action of the JPA is hence characterized by a mean output photon number
\begin{equation}
N=\frac{G^2+G^{-2}}{2} \overline{n} + \frac{G^2+G^{-2}-2}{4},
\end{equation}
where $\overline{n}$ is the input mean photon number (${\cal{G}}\equiv \frac{G^2+G^{-2}}{2}$ is called the \textit{power gain}). The two-photon coherence,
\begin{equation}
M=\frac{G^2-G^{-2}}{4} \left(2\overline{n}+1\right)
\end{equation}
characterizes the degree of squeezing by the phase sensitive amplification. In the case of a vacuum input state, the mean output photon number is
$N=\frac{G^2+G^{-2}-2}{4}$ and we have $M=\sqrt{N(N+1)}$.

If the JPA driving flux modulation at $2\omega_0$ has a phase $2\phi$, the squeezing occurs for a rotated quadrature component and is represented by a complex value, $M\rightarrow M e^{2 i\phi}$ in Eq. ~\ref{eq:field-temp}. Using the expressions in Eq.~\ref{eq:field-temp}, we readily find that a rotated quadrature component $\hat{X}_\theta = \frac{1}{2}(e^{-i\theta} \hat{a}_{\mathrm{out}}+e^{i\theta} \hat{a}_{\mathrm{out}}^\dagger)$ of the squeezed state has a variance,
\begin{equation} \label{eq:theta}
\delta X_\theta^2= \langle \hat{X}_\theta^2\rangle =  \frac{1}{2}\left[N + M\cos\left(2(\theta-\phi)\right) + \frac{1}{2}\right].
\end{equation}
The angles $\theta=\phi$ and $\theta=\pi/2+\phi$ specify the principal axes of the squeezing ellipse, along which the fluctuations are anti-squeezed and squeezed by the phase sensitive gain factors $G$ and $G^{-1}$, respectively.

In the case of steady state squeezing with a finite bandwidth, $\Delta_{sq}$, the delta-function correlations in Eqs.~\ref{eq:field-temp} are replaced by exponential functions in the time argument $|t-t'|$ ~\cite{Gardiner.QuantumNoise}. In Section II of the main text, the squeezed output field $\hat{a}_{\mathrm{out}}$ is taken as the input to the resonator system containing the probed spin ensemble.

\subsection{Resonator output signal and its fluctuations}

We now present details of the derivation of the theoretical expressions given in Section II of the main text for the amplitude and noise properties of the cavity output field. We provide closed-form analytical results in the special case of a Lorentzian spin-frequency distribution uncorrelated with the coupling strengths.

Applying the Fourier transforms, $\tilde{h}(\omega) = \frac{1}{\sqrt{2\pi}}\int h(t) e^{-i\omega t}\, dt$ and $h(t) = \frac{1}{\sqrt{2\pi}}\int  \tilde{h}(\omega)e^{i\omega t} \,d\omega$, the equations of motion Eqs.\,(2, 3) in \cite{mainText} can be written in algebraic form,
\begin{align}
\begin{split}
-i\omega\tilde{a}(\omega) &= -\frac{\kappa}{2}\tilde{a}(\omega) -i \sum_j g_j \tilde{\sigma}_j(\omega)
+ \sqrt{\kappa_L}\tilde{b}_{\mathrm{loss}}(\omega)
+\sqrt{\kappa_C} \tilde{b}_{\mathrm{in}}(\omega)
\\
-i\omega \tilde{\sigma}_j(\omega) &= -(\gamma+i\Delta_j)\tilde{\sigma}_j(\omega) - ig_j \tilde{a}(\omega) + \frac{\alpha}{\sqrt{2\pi}} e^{i\Delta_j \tau} + \sqrt{2\gamma}\tilde{F}_j(\omega).
\end{split}
\end{align}
The equations for the spin operators can be formally solved and subsequently yield the expression for the intra-cavity field operator
\begin{equation} \label{eq:fieldfinal}
\tilde{a}(\omega) = \frac{-\frac{i}{\sqrt{2\pi}}\sum_j \alpha g_j e^{i\Delta_j \tau}/(\gamma+i\Delta_j-i\omega) +\tilde{F}_{\mathrm{spin}}(\omega)+\sqrt{\kappa_L} \tilde{b}_{\mathrm{loss}}(\omega)+ \sqrt{\kappa_C}\tilde{b}_{\mathrm{in}}(\omega)}{\kappa/2 - i\omega + \sum_j g_j^2/(\gamma+i\Delta_j-i\omega)}.
\end{equation}
In this equation, 
\begin{align}
\tilde{F}_{\mathrm{spin}}(\omega)=-i\sqrt{2\gamma}\sum_j g_j\frac{\tilde{F}_j(\omega)}{(\gamma+i\Delta_j-i\omega)}
\end{align}
with $[\tilde{F}_{\mathrm{spin}}(\omega),\tilde{F}_{\mathrm{spin}}^\dagger(\omega^\prime)] = 2\gamma \sum_j g_j^2 \frac{\delta(\omega-\omega^\prime)}{\gamma^2+(\Delta_j-\omega)^2}$,
accounts for the contribution of the spins to the noise in the cavity field.

The output field operator is related to the cavity mode operator by the input-output relation Eq.\,(4) in \cite{mainText}, which for a cavity field of the form Eq.~\eqref{eq:fieldfinal} allows the relative contributions of the spin ensemble and noise sources to be parametrized by a set of four complex frequency dependent coefficients as given in Eqs.\,(6-9) in \cite{mainText}. The result is displayed in Eq.\,(5) in \cite{mainText}, and we note that the output field is completely determined by the frequency dependent ensemble cooperativity Eq.\,(10) in \cite{mainText} and the amplitude factor Eq.\,(11) in \cite{mainText}.
In the following, we outline how a few realistic assumptions lead to analytic expressions for these quantities.

The spin Larmor frequencies have a given distribution $f(\Delta)$ and with a large number of spins, we may replace the sum over spins in $\tilde{a}(\omega)$ by an integral $\sum_j \boldsymbol{\cdot} \rightarrow N_{\mathrm{spins}}\int d\Delta\,f(\Delta) \boldsymbol{\cdot}$. For simplicity, we assume that the variation in the coupling strengths $g_j$ is small and uncorrelated with the frequency distribution, such that we may introduce a mean coupling constant $g=\sqrt{\frac{1}{N_{\mathrm{spins}}}\sum_j |g_j|^2}$.
The sum in the denominator of Eq.\,(\ref{eq:fieldfinal}) can then be written as $\kappa C(\omega)/2$ by introducing the frequency-dependent ensemble cooperativity,
\begin{equation}\label{eq:C}
C(\omega) = \int\frac{2g^2N_{\mathrm{spins}}f(\Delta)}{\kappa(\gamma+i\Delta-i\omega)}\, d\Delta.
\end{equation}
Likewise, the sum in the numerator can be written as
\begin{equation}\label{eq:A}
A(\omega) = \int \frac{g\alpha N_{\mathrm{spins}}f(\Delta) e^{i\Delta \tau}}{\gamma+i\Delta-i\omega}\,d\Delta.
\end{equation}
Finally, the magnitude of the spin noise contribution $\tilde{F}_{\mathrm{spin}}(\omega)$ in the numerator is similarly assessed by replacing the sum in the commutator relation by an integral yielding $[\tilde{F}_{\mathrm{spin}}(\omega),\tilde{F}_{\mathrm{spin}}^\dagger(\omega^\prime)] = \kappa \mathrm{Re}\left[C(\omega)\right]\delta(\omega-\omega^\prime)$.

From the expressions above, the cooperativity and amplitude factors may be evaluated for any distribution of spin frequencies.
As an illustrative example, we consider a Lorentzian lineshape of the spin Larmor frequencies, $f(\Delta) = (\Gamma/2\pi)/(\Delta^2 + \Gamma^2/4)$. The integrals in Eqs.\,(\ref{eq:C}, \ref{eq:A}) can then  be performed analytically. This yields the frequency dependence of the ensemble cooperativity,
\begin{equation}
C_{\mathrm{Lorentzian}}(\omega) = \frac{2 g^2N_{\mathrm{spins}}[\Gamma/2+\gamma+i\omega]}{\kappa [(\gamma+\Gamma/2)^2+\omega^2]}.
\end{equation}
The numerator of the integrand in $A(\omega)$ has complex poles at $\Delta=i\Gamma/2 + \omega$ and $\Delta=i\gamma + \omega$. Due to the the exponential factor $e^{-\Gamma\tau/2}$ evaluated at the echo time which occurs much later than $1/\Gamma$, the contribution from the first pole is negligible. The second pole, however, yields the weaker damping factor $e^{-\gamma\tau}$ and a finite contribution,
\begin{equation}
A_{\mathrm{Lorentzian}}(\omega)=(\alpha/\sqrt{2\pi}) (N_{\mathrm{spins}}\Gamma/[(i\gamma+\omega)^2+\Gamma^2/4])e^{-\gamma\tau+i\omega\tau}
\end{equation}
with a frequency width $\sim \Gamma/2$.

\section{Experimental details}

\subsection{Measurement setup}
\begin{figure}[htbp!]
\includegraphics[width=\linewidth]{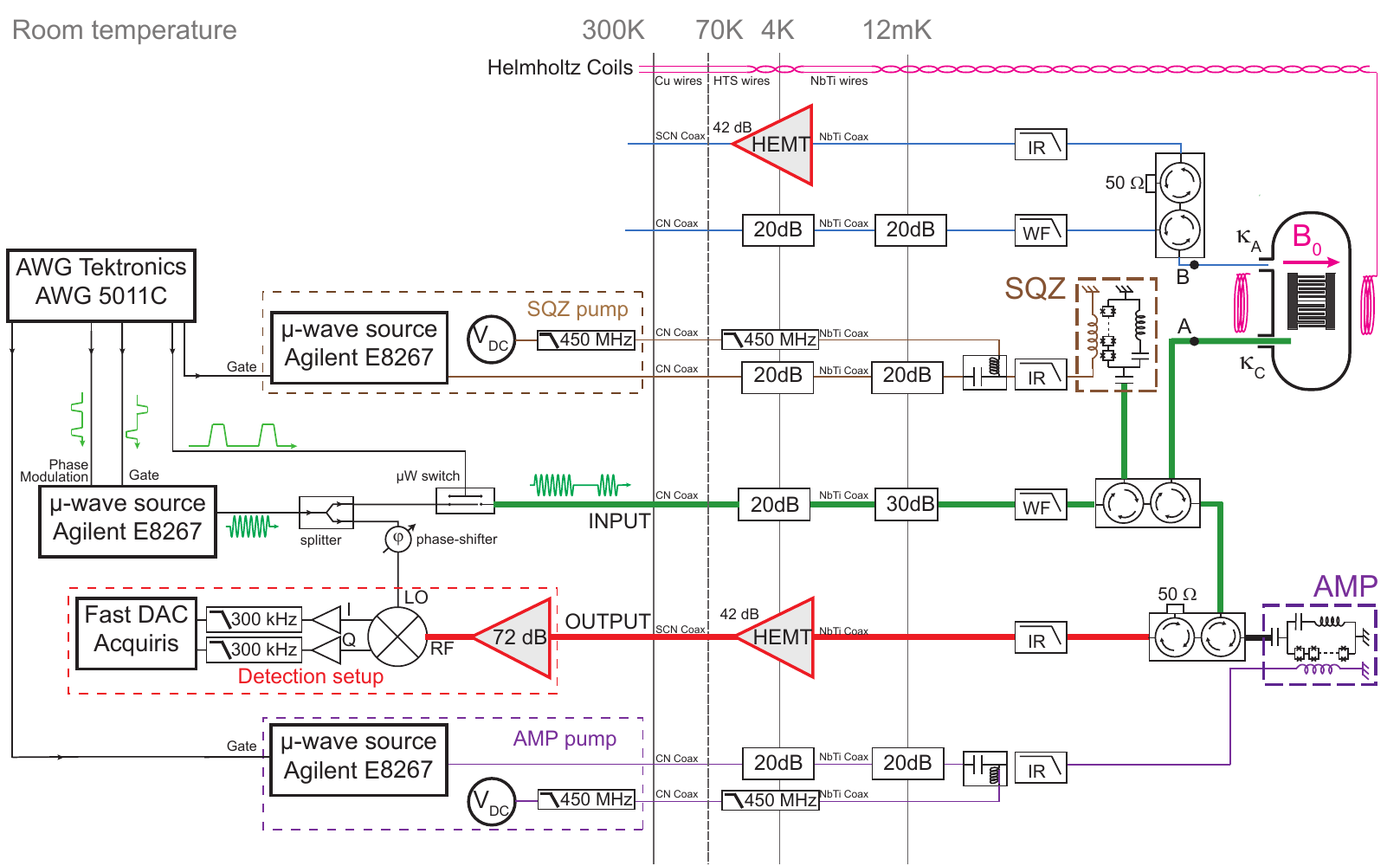}
\caption{\textbf{Measurement setup} \label{figSetup}}
\end{figure}
\ay{ The detailed microwave setup is shown in Fig.~\ref{figSetup}. The experiment described in the main text and schematized in Fig.~2c makes use of the green and red lines. They connect successively the SQZ, the ESR resonator coupled at rate $\kappa_C$, the AMP, followed by the HEMT and room-temperature amplification. The green input port is used to transmit the spin driving pulses to the ESR resonator. The ESR resonator design and fabrication details are given in \cite{bienfait2015reaching}.}

\ay{In addition, the setup includes additional input-output lines to probe the LC resonator in reflection on port 1 (blue lines), in transmission (blue-green lines), and in reflection on port 2 (green-red lines). Measurement of the full resonator scattering matrix and a fit to the resonator input-output formulas~\cite{PalaciosLaloy2010} yields $\kappa_A= 3 \cdot 10^3 ~ \mathrm{s}^{-1}$, $\kappa_C= 1.6 \cdot 10^6 ~\mathrm{s}^{-1}$ and $\kappa_{L}=6 \cdot 10^4 ~\mathrm{s}^{-1}$.}

Input lines are isolated from microwave photons emitted from higher temperature stages by a minimum of $20$~dB at $4$~K and $20$~dB at $20$~mK, and from infrared photons by commercial absorptive filters (Wavefade FLP0960) anchored at $20$\,mK. Output lines are protected from microwave noise by a minimum of two circulators and and from infra-red photons by home-made absorptive filters. The ESR resonator and both JPAs are magnetically shielded, see \cite{bienfait2015reaching} for details.


The generation of the microwave pulses at the input and the detection setup are as described in Ref.\,\cite{bienfait2015reaching}. As explained in the main text, AMP and SQZ have the same design~\cite{Zhou.PhysRevB.89.214517(2014)}, and both devices can be tuned to the desired operating frequency via a DC bias of the flux threading their SQUID-array. The two microwave pump tones are generated by microwave sources locked with a $1$\,$G$Hz synchronization loop to the microwave source providing the excitation pulses and the local-oscillator tone to ensure phase stability. The pump  tones are in addition pulsed via the microwave source internal switches to generate gain only when needed.

\subsection{Squeezing-enhanced echo: data acquisition and processing}

\begin{figure}[htbp!]
\includegraphics[width=\linewidth]{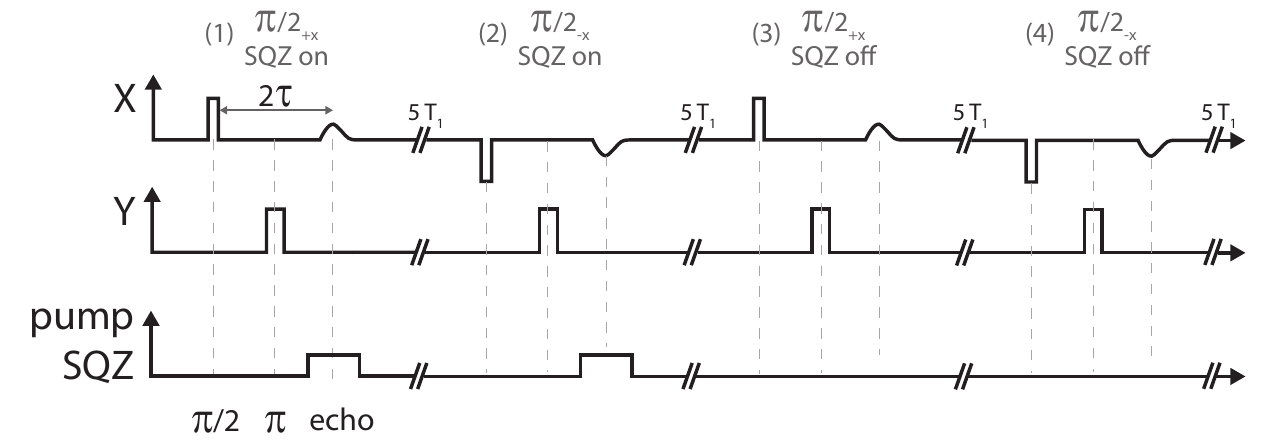}
\caption{\textbf{Experimental acquisition sequence}\label{figSeq} taken with $\tau=200~\upmu$s. Phase cycling as well as alternate SQZ switching are used to compensate setup drifts during the $6$~h long acquisiton.}
\end{figure}

We describe in the following the acquisition and post-processing of the echo data shown in \ay{ Fig.~5(c and d)} of the main text. To minimize setup drifts, we alternate echos acquired with SQZ on and with SQZ off as well as use phase-cycling, as shown in Fig.~\ref{figSeq}. We acquire $N_{\mathrm{avg}}=2500$ echos with SQZ on and 2500 SQZ off. The quadrature voltage $I(t)$ is digitalized at a sampling rate of $1 \rm{pt}/\rm{\mu s}$ with an acquisition bandwidth of $300$~kHz. The data is recorded in a time window $T=70\,\rm{\upmu s}$ centered on the echo. The waiting time between each echo sequence is taken to be $T_{\mathrm{rep}}\approx 5 T_{\rm 1}=5$\,s.

We compute the averaged signals shown in \ay{  Fig.~5b} of the main text as:
\begin{equation}
 \bar{I}_{\mathrm{on}} (t) = \sum^{N_{\mathrm{avg}}}_{i = 1} \frac{I_{(1), i}
   (t) - I_{(2), i} (t)}{2}\:  {\rm{and}}\: \bar{I}_{\mathrm{off}} (t) =
   \sum^{N_{\mathrm{avg}}}_{i = 1} \frac{I_{(3), i}  (t) - I_{(4), i} (t)}{2}
\end{equation}
where subscripts $(i)$ are denoted in Fig.~\ref{figSeq}. The noise histograms in Fig.~3c are computed from the bins $\lbrace I_{(1), i}(t)- \bar{I}_{\mathrm{on}}(t), \forall i, \forall t\rbrace \cup \lbrace I_{(2), i}(t)+ \bar{I}_{\mathrm{on}}(t), \forall i, \forall t\rbrace$ when the SQZ is on and $ \lbrace I_{(3), i}(t)- \bar{I}_{\mathrm{off}}(t), \forall i ,\forall t\rbrace \cup \lbrace I_{(4), i}(t)+ \bar{I}_{\mathrm{off}}(t), \forall i ,\forall t\rbrace$. To ensure the echo emission is not affecting the noise properties, we have also computed the noise histograms and variances keeping only identical stamping times $t$ and found no variations.

To compute the SNR for both echos, we define modes of the propagating field as $\langle O\rangle =\frac{1}{T} \langle O(t)\rangle u(t) dt $ with $u(t)$ a filter function normalized as $\frac{1}{T} \int [u(t)]^2 dt=1$ \cite{bienfait2015reaching}. We pick as filter function the echo averaged signal $u(t)\propto\left[{I}_{\mathrm{on}} (t)+{I}_{\mathrm{off}} (t)\right]/2$. For each echo $\lbrace (n),i \rbrace$, we can thus evaluate the signal and noise quantities as $\langle I_{(n),i} \rangle$ and $\sqrt{\langle \Delta I_{(n),i}^2 \rangle}$. Averaging over all recorded echos yields the noise and echo signal shown in Table~\ref{tableResults}, demonstrating a noise reduction by $11\%$. Repeating the same procedure for a tophat $u$ function  of width $20~\upmu$s centered on the echo yields similar results.

\begin{table}
\setlength{\tabcolsep}{10pt}

\begin{tabular}{|c|c|c|c|c|c|c|}
\hline
$u(t)$ & \multicolumn{3}{c|}{Echo shape} & \multicolumn{3}{c|}{Top Hat function} \\
\hline
SQZ &  $\langle I \rangle$ & $\sqrt{\langle \Delta I^2 \rangle}$ & SNR & $\langle I \rangle$ & $\sqrt{\langle \Delta I^2 \rangle}$ &SNR \\
\hline
OFF & 0.179 & 0.202 & 0.886  & 0.161 & 0.202 & 0.797 \\
\hline
ON & 0.177 & 0.181 & 0.973 & 0.160 & 0.181 & 0.884 \\
\hline
ON/OFF ratio & 0.988 & 0.897 &  1.10 & 0.992 & 0.894 & 1.11 \\
\hline
\end{tabular}
\caption{\textbf{Experimental results}\label{tableResults}.}
\end{table}

\subsection{Sensitivity estimate and numerical model}\label{simus}

To estimate the sensitivity of the experiment and its improvement when using squeezed vacuum, we shall determine the number of spins contributing to the echo signal shown in Fig.~5 of the main text. This number is defined as the number of spins excited by the first $\pi/2$ pulse of the Hahn echo sequence. The resonator bandwidth is 20 times smaller than the spin linewidth, broadened due to induced strain, and as in \cite{bienfait2015reaching} we thus need to resort to numerical simulations. \ay{In these simulations, the system is modelled as $N_{D}$ spin-$1/2$ systems coupled to the resonator following Eq.\,(1) in \cite{mainText}.} The evolution of the spin observables and the intra-resonator field mean-values is then found by integrating the master equation of the system  with the resonator leakage and the spin decoherence taken into account in a Born-Markov approximation \cite{Grezes.PHYSREVX4.021049(2014)}. To describe the inhomogeneity in spin frequencies and coupling constants, the spin ensemble is divided in $M$ sub-ensembles with coupling constants $g^{(m)}$ and detunings from the resonator frequency $\Delta^{(m)}$.

In our former work \cite{bienfait2015reaching}, using the same sample and a resonator of similar geometry, additional measurements such as the absorption of a weak microwave pulse and Rabi oscillations allowed us to determine the spin distribution at $B_0=2.8$~mT (peak of the line) to be well modelled by a Gaussian distribution in $g$ of central value $g_0/2\pi=56$~Hz and FWHM $\Delta g=1.5$~Hz and a square distribution for the spin detunings $\Delta$ with a width far exceeding the resonator bandwidth. Both distributions were weighted with a total overall factor $N_D=3.6\times 10^5$. Compared to \cite{bienfait2015reaching}, the resonator presented in the main text has a 10 times lower quality factor, corresponding to a 10 times larger linewidth and damping rate. Repeating the numerical simulations of \cite{bienfait2015reaching}, taking into account these modifications, we characterize the number of spins contributing to the signal from the $B_0=2.8$~mT peak to be $N_{\mathrm{spins}}=1.2\times10^5$.

\begin{figure}[htbp!]
\includegraphics{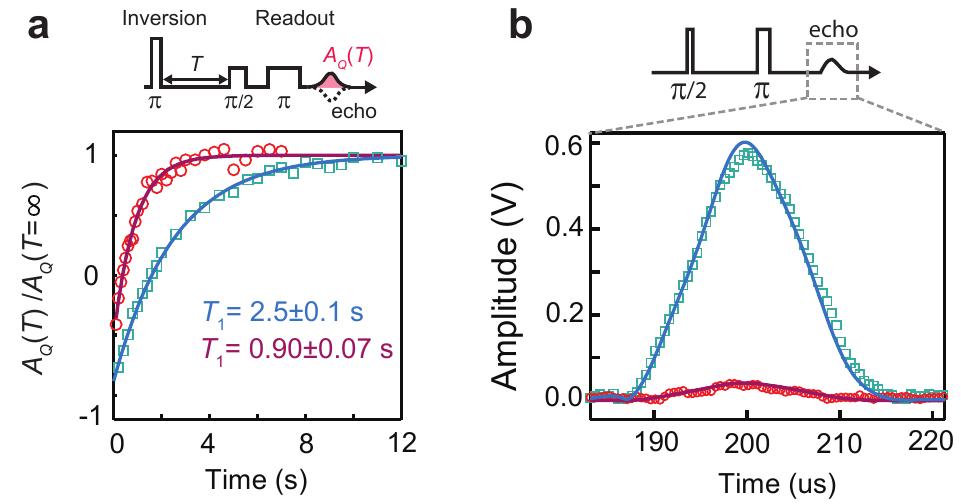}
\caption{\textbf{Calibration of number of excited spins.} In both panels, data acquired at $B_0=2.6$~mT ($B_0=2.8$~mT) are represented by red circles (blue squares) \textbf{a} Spin relaxation times measured using inversion recovery, well fit by exponential decays (solid lines). \textbf{b} Echo time traces reproduced via numerical simulations (solid lines).\label{figNSpins}}
\end{figure}

\ay{ We now need to characterize the number of spins at the magnetic field used in the main text, $B_0=2.6$~mT. We would like to proceed by comparing the echo amplitudes recorded at $B_0=2.6$~mT and $B_0=2.8$~mT. We find experimentally a ratio $\alpha\approx \times 25$. However, a direct comparison is not possible since these two spin subsets can not be modelled by the same coupling constant distribution.} Indeed, the spin resonance frequency distribution is caused by strain applied by thermal contraction of the aluminium on the silicon substrate. As a result, the spin spectral and spatial distributions are linked~\cite{pla2016}. Spins on the low-field side of the peak at $B_0=2.6$~mT correspond to spins lying near the edge of the aluminium wire, whereas spins contributing to the $B_0=2.8$~mT peak correspond to spins located under the central part of the wire. Since the aluminium wire is superconducting, the current density is higher on the edge of the wire than in the central part, and spins at $B_0=2.6$~mT are thus more strongly coupled to the resonator than spins at $B_0=2.8$~mT. To estimate the difference in $g$, we measure the spin relaxation time $T_{\rm 1}$. As $T_{\rm 1}$ is radiatively limited by the Purcell effect in our experiment \cite{bienfait2016Purcell}, we have $T_{\rm 1}^{-1}=4g^2/\kappa$. Measuring $T_{\rm 1}$ at $B_0=2.6$~mT and $B_0=2.8$~mT (see Fig.~\ref{figNSpins}a) and assuming that  only the central value $g_0$ of the spin distribution should be adjusted, we determine $g_{0}^{(2.6 ~{\rm mT})}/2\pi=93$~Hz. 

We next adjust the distributions overall weighting factor $N_D$ until we are able to reproduce the difference in the echo amplitudes recorded  at $B_0=2.6$~mT and $B_0=2.8$~mT (see Fig.~\ref{figNSpins}b). We find $N_D=1.4\times 10^4$ and thus infer the number of excited spins to be $N_{\mathrm{spins}}=4.7\times 10^3$.

Finally, we determine the sensitivity $N_{\mathrm{min}}$ defined in the main text as the minimum number of spins detectable per echo with a SNR of 1. In the data shown in Fig.~5c of the main text, the single-shot SNR is $0.37$ in the absence of squeezing and we hence have $N_{\mathrm{min}}=N_{\mathrm{spins}}/SNR=1.3\times10^4$. This larger value compared to~\cite{bienfait2015reaching} is due to the lower quality factor of the resonator.

Thanks to these numerical simulations, we can also check the assumptions of low cooperativity made in Section~II of the main text. Using the model corresponding to $B_0=2.6$~mT, we find $\ay{ C(0)}=0.002$, verifying $\ay{ C(0)}\ll 1$ and thus confirming that the spin-noise contribution can be neglected at low squeezing. 

\section{Characterization of the mean number of thermal photons}

In order to produce a squeezed vacuum state of the field $\hat{b}_{\mathrm{in}}$ that serves as input to the ESR resonator, the squeezer "SQZ" is pumped while its input is in the vacuum state (see Fig.~\ref{figSetup}).
However, due to imperfect filtering of the microwave probe lines and to the refrigerator finite base temperature, one can never  reach perfect electromagnetic vacuum. In this section, we describe the calibration procedure used to place an upper bound  on the average excitation number of the input Fourier modes $\tilde{b}_{\mathrm{in}}(\omega) $ around the ESR resonator resonance frequency  (typically $|\omega-\omega_0| \leqslant \kappa_c$)  when the squeezer is off.  Note that this average excitation number is in fact characterized at a slightly different frequency $\omega_1/2\pi = 7.62~\mathrm{GHz}$, but we assume that the thermal equilibrium is similar so that $\langle \tilde{b}^{\dagger}_{\mathrm{in}}(\omega)\tilde{b}_{\mathrm{in}}(\omega) \rangle=\langle \tilde{b}^{\dagger}_{\mathrm{in}}(\omega_1)\tilde{b}_{\mathrm{in}}(\omega_1) \rangle$ for all relevant values of $\omega$. This assumption is reasonable given that $|\omega-\omega_1| \ll k_B T$ ($T$ being either the refrigerator base temperature, or the effective temperature of the modes given at the end of this section), and that the transmission of the microwave input lines is flat ($\pm 0.5~\mathrm{dB}$) variation) on this frequency range.

The method consists in replacing, in a subsequent cool-down of the refrigerator, the ESR resonator with  a transmon superconducting qubit \cite{koch2007charge}   coupled to a microwave readout resonator with resonance frequency $\omega_1$. 
The resonator-qubit system is in the so-called \textit{strong dispersive regime} of circuit QED in which photons in the resonator mode lead to dephasing of the qubit~\cite{gambetta2006qubit,blais2007quantum}. Thus, by measuring the  dephasing rate of the qubit beyond the effect of population relaxation $\gamma_{\phi}=\gamma_2-\gamma_1/2$, one can   place an upper bound on the thermal photon number in the readout resonator, and then on the occupation of the travelling modes $\hat{b}_{\mathrm{in}}$.


\begin{figure}[h]
\includegraphics[width=0.7\columnwidth]{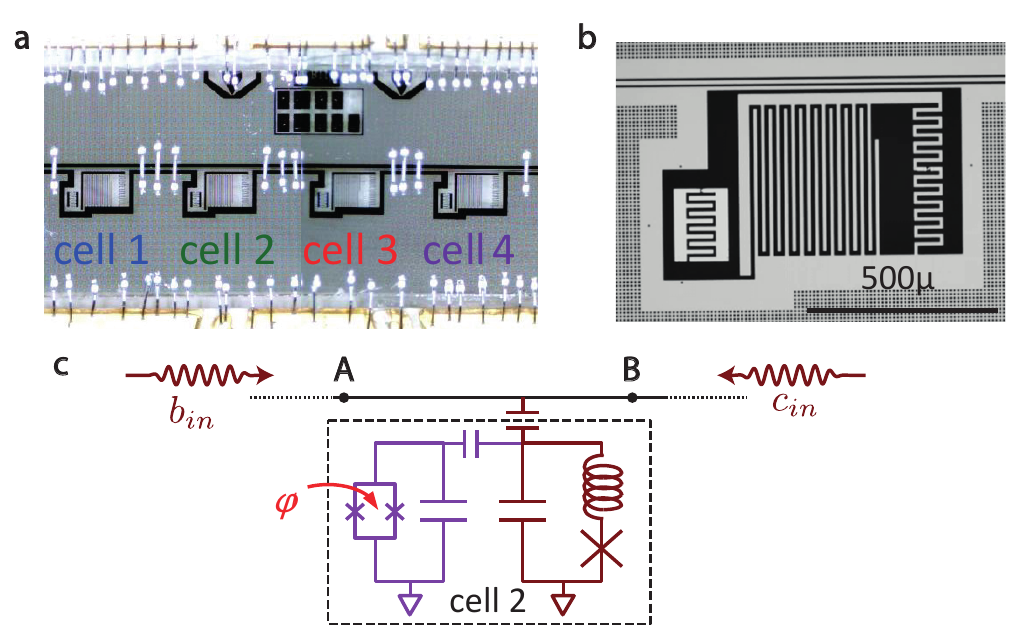}
\caption{\textbf{a)} Optical micrograph of the device used for estimating the number of thermally excited photons in $\hat{b}_{\mathrm{in}}$.  Four cells, each one composed of a  transmon qubit with an attached readout resonator are probed with a single microwave feedline. \textbf{b)} Zoom on one of the cells, showing the tunable transmon qubit (to the left) capacitively coupled to the readout resonator (to the right), itself capacitively coupled to the feedline (to the top). \textbf{c)} Simplified  electric circuit of the cell used for the calibration. The qubit contains a split Josephson junction and its resonance  frequency can be tuned  by threading the loop with a magnetic flux $\varphi$. The resonator, which contains an array of junctions, is slightly non-linear. Thermal excitations in the resonator are due to both right and left travelling modes $\hat{b}_{\mathrm{in}}$ and $\hat{c}_{\mathrm{in}}$ in the feed line.
\label{figS6}}
\end{figure}

The device that we use was not designed specifically for the experiment, but was studied in Ref.~\cite{schmitt2014multiplexed}. On a sapphire chip, 4 lumped element microwave readout resonators, each one capacitively coupled to a transmon qubit (see Fig.~\ref{figS6}), are  coupled to a single transmission feed line.  In the following, we consider only the qubit-resonator system labeled \textit{cell 2} (the other ones are well out of resonance). The feed line is connected to the setup depicted on Fig.~S1 at points A and B. Note that in this geometry, the readout resonator thermal occupation is set by the average occupation of right propagating modes $\hat{b}_{\mathrm{in}}$ through A  and left propagating modes $\hat{c}_{\mathrm{in}}$ through B (see Fig.~\ref{figS6}~a). Internal losses of the readout resonator, that could act as a coupling to a fictitious cold reservoir, are shown to be negligible on Fig.~\ref{figS8}~d. The blue input  line connected at B on Fig.~\ref{figSetup}, which was originally designed to probe the ESR resonator in reflection on port 1, is  less attenuated by 10 dB than the green line connected to A so that left propagating modes tend to increase the thermal occupation of the readout resonator. This issue does not arise with the ESR resonator since the coupling rate through port 1 is negligible ($\kappa_1 \ll \kappa_2$). Thus, the calibration made here is conservative and the estimation of the thermal occupation of $\hat{b}_{\mathrm{in}}$ is an upper bound of the actual value in the experiment.

\begin{figure}[h!]
\includegraphics[width=0.6\columnwidth]{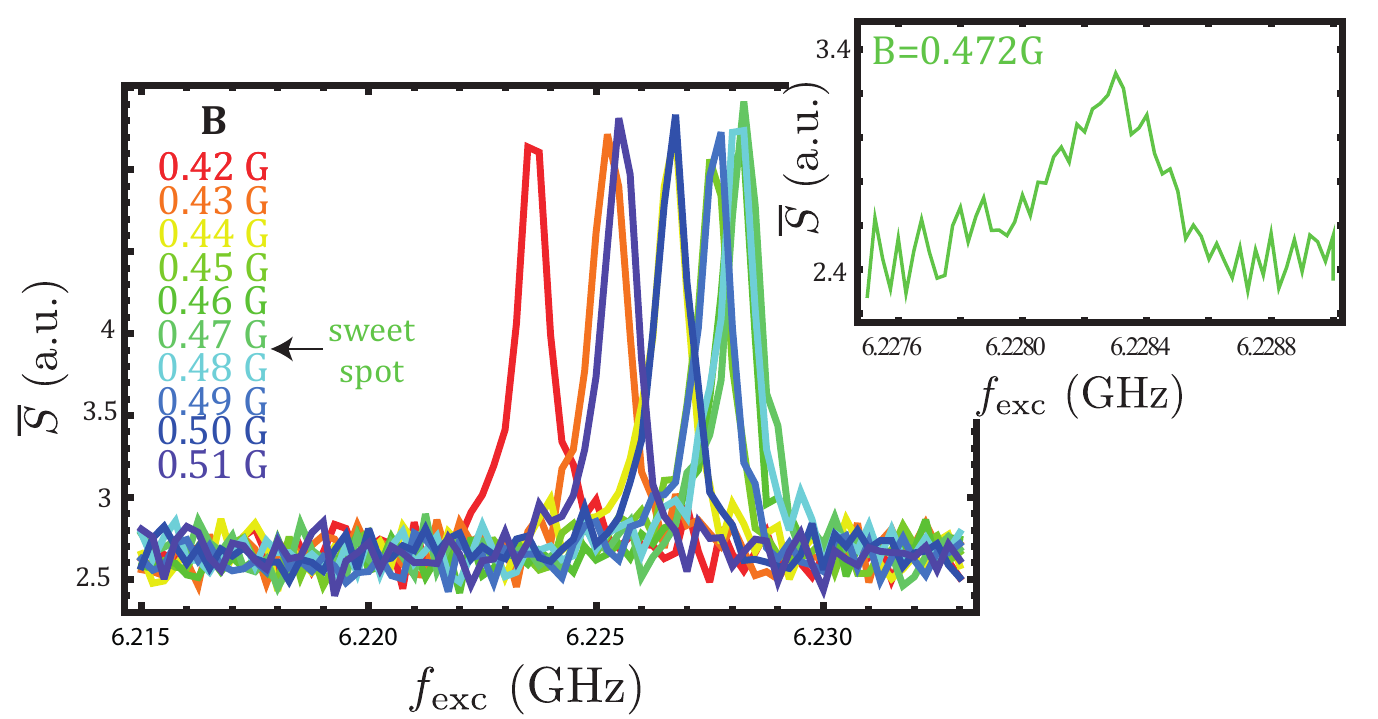}
\caption{\textbf{Two-tone spectroscopy} of the qubit. Starting from thermal equilibrium, the qubit is excited by a $5~\mu\mathrm{s}$-long saturating pulse (power -20~dBm referenced at refrigerator input) of frequency $f_{exc}$ and then readout with an optimized pulse around $\omega_1/2\pi$ (see text and Fig.~\ref{figS8}). The integrated signal $\overline{S}$ reveals the qubit excited state occupation. One can vary the qubit resonance frequency by varying the amplitude of the applied B-field (encoded in color). \textbf{Inset:} desaturated qubit resonance (power -30~dBm at fridge input) at the sweet spot, showing that $\omega_q/2\pi=6.228~\mathrm{GHz}$. \label{figS7}}
\end{figure}

\begin{figure}[h]
\includegraphics[width=0.75\columnwidth]{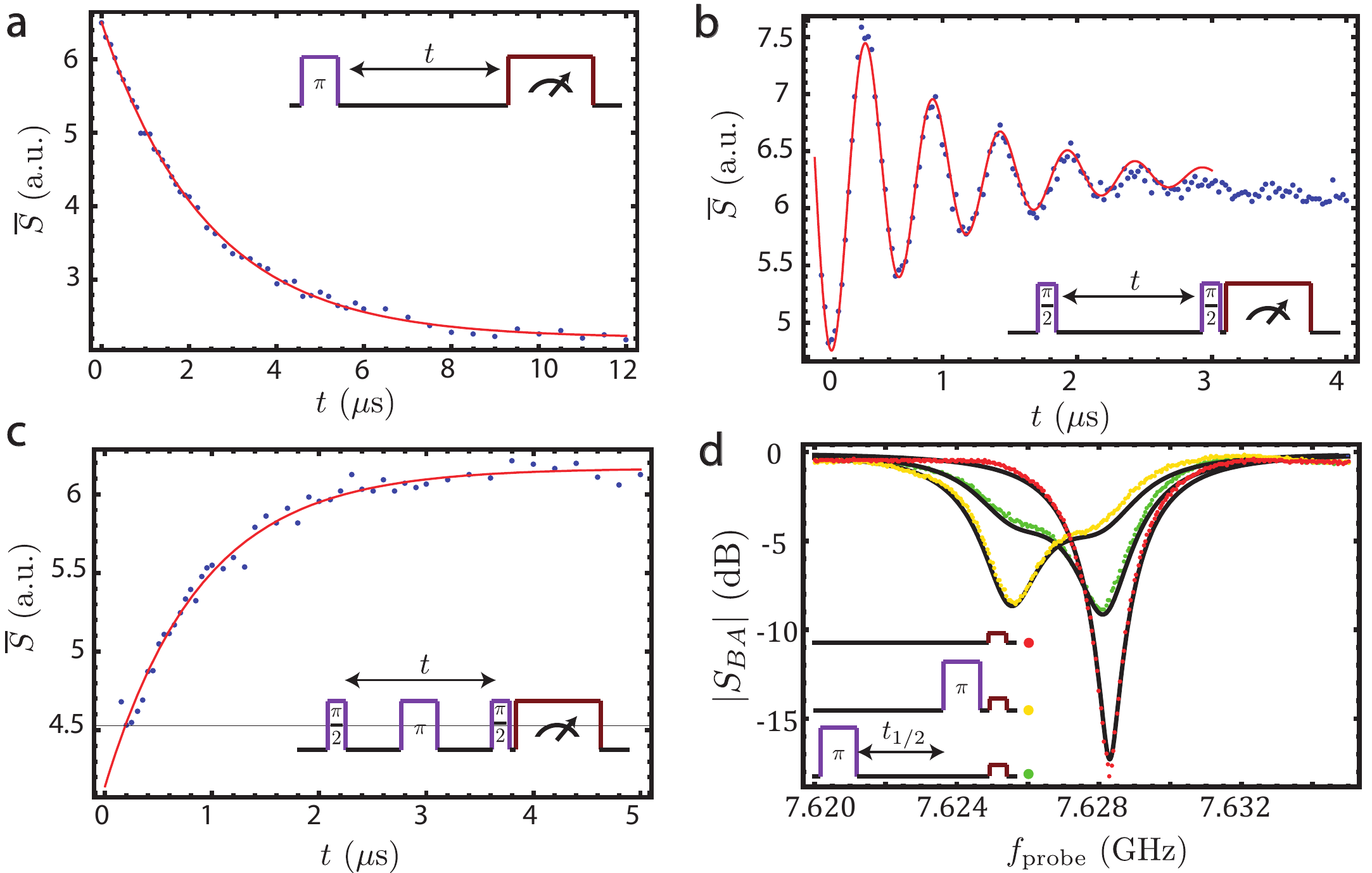}
\caption{\textbf{Qubit-resonator parameters characterization}. For each measurement, the pulse sequence is schematically represented at $\omega_q$ (in purple, all rotations around $\sigma_y$ of the qubit) and at $\omega_{\mathrm{readout}}\simeq \omega_1$ (in brown). \textbf{a)} Population relaxation measurement yielding $T_1=2.4~\mu\mathrm{s}$, \textbf{b})  Free induction decay measurement yielding $T_2^{\ast}=0.9~\mu\mathrm{s}$ (excitation pulses at $\omega_q/2\pi+2~\mathrm{MHz}$). \textbf{c)}  Hahn-echo measurement yielding $T_{2, \mathrm{echo}}=0.9~\mu\mathrm{s}$. \textbf{d)} Measured transmission coefficient $S_{AB}$ when the qubit is at thermal equilibrium (red dots), right after an inverting $\pi$-pulse (yellow dots) and a qubit half-life after a  $\pi$-pulse (green dots). Black lines: global fit with parameters $p_{\pi}=0.66$, $\chi/2\pi=1.48~\mathrm{MHz}$ and $\kappa_{\mathrm{int}}/\kappa_{\mathrm{ext}}=0.14$. For \textbf{a}, \textbf{b} and \textbf{c} the readout pulse power is empirically adjusted to optimize signal to noise ratio and the transmitted field is integrated over $5~\mu\mathrm{s}$. Only the quadrature $\overline{S}$ containing information on the qubit state is plotted. For \textbf{d} the readout pulse power is low enough that readout resonator non-linearity is neglected and the transmitted field is integrated over $0.2~\mu\mathrm{s}$ in the stationary regime.
\label{figS8}}
\end{figure}

The readout resonator consists of an interdigitated capacitor made out of superconducting aluminum in parallel with an array of Josephson junctions (Fig.\,\ref{figS6}). This array behaves as a  non-linear inductor and was originally designed for single-shot readout of the attached qubit. This non-linearity is not relevant here and can be neglected as the average photon number in the resonator is well below one. The transmon qubit is made out of a smaller interdigitated capacitor in parallel with a split Josephson junction that allows to tune its resonance frequency. A DC magnetic field is then applied using a superconducting coil in order to operate the device at its \textit{sweet spot}, where its frequency $\omega_q/2\pi =6.23~\mathrm{GHz}$ does not depend on the magnetic field fluctuations to first order (see Fig.~\ref{figS7}).

The coupling rate of the qubit and readout resonator is much smaller than the detuning $\omega_1-\omega_q$ so the system is described by the dispersive hamiltonian~\cite{blais2007quantum}
 \begin{equation}
 \hat{H}=\hbar \omega_{\mathrm{1}} (\hat{a}^{\dagger} \hat{a}+ \frac{1}{2})+ \hbar \omega_q \frac{\hat{\sigma}_z}{2} +\hbar \chi \hat{a}^{\dagger} \hat{a} \hat{\sigma}_z.
 \end{equation}
Here, $\hat{\sigma}_z$ is the Pauli operator of the qubit and $\chi$ is the qubit state dependent shift of the readout resonator frequency, which provides us with a robust readout method of the transmon \cite{wallraff2005approaching,reed2010high}. Indeed, by probing the  resonator with a near resonant microwave field and integrating a quadrature of the transmitted field, one gets a signal $\overline{S}$ depending linearly on $\langle \sigma_z \rangle$.  In practice, the power, duration and frequency of the readout pulse was empirically adjusted to optimize signal-to-noise ratio. It corresponds to few photons in the resonator (power 10~dB larger than for 1~photon characterization of the resonator on Fig~\ref{figS8}~d). Note that the amplifier JPA was turned off during all measurements.

In Ref.\,\cite{rigetti2012superconducting},  Rigetti \textit{et al.} computed the dephasing rate of a qubit induced by thermally excited photons in the readout resonator mode. It reads
\begin{equation}
\gamma_{\mathrm{phot}}=\frac{\kappa}{2}\mathrm{Re} \left\{ \sqrt{\left(1+2i\frac{\chi}{\kappa}\right)^2+8i\overline{n}\frac{\chi}{\kappa}} -1\right\},
\label{eqdeph}
\end{equation}
where $\kappa$ is the photon exit rate from the readout resonator and $\overline{n}=\langle \hat{a}^{\dagger}\hat{a}\rangle$ is the mean number of photons hosted by the resonator. Considering that $ \gamma_{\mathrm{phot}} \leq \gamma_{\phi} =\gamma_2 - \gamma_1/2$, we now measure the qubit population and coherence relaxation rates  $\gamma_1$ and $\gamma_2$ as well as all parameters entering the expression~\ref{eqdeph} in order to place an upper bound on $\overline{n}$.

By applying $\pi$ and $\pi/2$ excitation pulses (calibrated by recording Rabi oscillations of the qubit), we first measure the qubit population relaxation rate $\gamma_1=0.41~\mu\mathrm{s}^{-1}$ (see Fig.~\ref{figS8}~a) and coherence relaxation rate $\gamma_2^{\ast}=1.1~\mu\mathrm{s}^{-1}$ (see Fig.~\ref{figS8}~b).   This last rate corresponds to a free-induction decay measurement, and includes the effect of low-frequency noise, such as second order effects of the fluctuations in the flux threading the qubit loop, along with high-frequency noise induced by thermal photons in the readout resonator. A Hahn-echo measurement, yielding a decay rate $\gamma_{2,\mathrm{echo}} \simeq \gamma_2^{\ast}$ shows that the former is negligible compared to the latter (see Fig.~\ref{figS8}~c). We can then extract the qubit pure dephasing rate $\gamma_{\phi}=\gamma_2 - \gamma_1/2=0.9~\mu\mathrm{s}^{-1}$.

In order to measure $\chi$, we then detect the transmitted signal through the feed line  for a probe pulse of low amplitude (linear regime of the readout resonator) and integrate the signal over  $0.2~\mu\mathrm{s} \ll T_1$ in the stationary regime of the resonator (signal during ring-up is discarded in order to avoid distortion of the signal). The transmission coefficient from A to B then reads \cite{pozar2009microwave}
 \begin{equation}
 S_{BA}(\omega)= p \frac{\kappa_{\mathrm{int}}+2i (\omega-\omega_{\mathrm{res}}-\chi)}{\kappa_{\mathrm{int}}+\kappa_{\mathrm{ext}}+2i (\omega-\omega_{\mathrm{res}}-\chi)}+ (1-p)\frac{\kappa_{\mathrm{int}}+2i (\omega-\omega_{\mathrm{res}}+\chi)}{\kappa_{\mathrm{int}}+\kappa_{\mathrm{ext}}+2i (\omega-\omega_{\mathrm{res}}+\chi)},
 \label{eq:transmit}
 \end{equation}
where $\kappa_{\mathrm{ext}}$ (resp. $\kappa_{\mathrm{int}}$) is the resonator photon exit rate into the feed line (resp. due to internal losses) and $p=\langle1- \hat{\sigma}_z \rangle$/2 is the occupation of the ground state of the qubit. Note that the total photon exit rate from the resonator $\kappa=\kappa_{\mathrm{int}}+\kappa_{\mathrm{ext}}=2.04\times10^7\rm{s}^{-1}$ is determined independently by measuring the ringdown time of the resonator.\\
We record this transmission coefficient at three different points: 
i) at thermal equilibrium ($p\simeq 1$, red dots on Fig.~\ref{figS8}~d), ii) right after applying an inverting $\pi$-pulse ($p=p_{\pi}$, yellow dots) and, iii) for better precision, a duration $t_{1/2}=\mathrm{ln}(2)T_1$ after a $\pi$-pulse ($p=p_{\pi}/2$, green dots). If $p_{\pi}$ can be roughly estimated given the drive pulse duration and delay before signal integration, it is difficult to predict accurately its value due to the reduction of $T_1$ in presence of a field in the readout resonator \cite{boissonneault2009dispersive}. We rather estimate it along with the other parameters entering Eq.~\ref{eq:transmit} by fitting these three curves altogether (black curves), which yields $p_{\pi}=0.66$,  $\chi/2\pi=1.48~\mathrm{MHz}$ and  $\kappa_{\mathrm{int}}/\kappa_{\mathrm{ext}}=0.14$. In this fit, we allow for a global scaling factor accounting for the attenuation in the lines, and  a small offset in the transmitted field complex amplitude, attributed to impedance mismatch.

From this calibration and using Eq.~\ref{eqdeph}, we find for the readout resonator $\overline{n}\leq 0.1$. As mentioned in the beginning of this section, it is a conservative estimate of the average thermal photon number in the ESR resonator mode when the squeezer is off. 

For a thermal state of a harmonic oscillator inside the ESR resonator when the squeezer is off, the fluctuations on one quadrature read
\begin{equation}\label{eq:thermal}
\delta X_{\mathrm{off}}^2  = \frac{\overline{n}}{2}+\frac{1}{4}.
\end{equation}
In the experiment, as seen in Figs.~3b,c of the main text, the detected output noise is reduced by a factor $\delta I^2_{\mathrm{on}} / \delta I^2_{\mathrm{off}}=0.75$ when the squeezer is on. Due to background noise in the transmission channels, this represents an upper bound on the reduction in the fluctuations of the squeezed quadrature inside the ESR resonator, $ \delta X^2_{\mathrm{on}}\ / \delta X^2_{\mathrm{off}}<0.75$. With a thermal occupation of less than 0.1 photons inside the resonator, Eq.\,(\ref{eq:thermal}) hence yields an upper bound on $\delta X^2_{\mathrm{on}}$,
\begin{align}
\delta X^2_{\mathrm{on}}<0.75 \left(\frac{\overline{n}}{2}+\frac{1}{4}\right) < 0.225,
\end{align}
showing that the fluctuations are indeed reduced below
the vacuum level ($\delta X^2_{\mathrm{vac}}=1/4$).

The background noise contribution can be estimated when both SQZ and AMP are switched off. This allows the statement of a more stringent condition $ \delta X^2_{\mathrm{on}}/  \delta X^2_{\mathrm{off}}<\frac{ \delta I^2_{\mathrm{on}}- \delta I^2_{\mathrm{bg}}}{\delta I^2_{\mathrm{off}}-\delta I^2_{\mathrm{bg}}}=0.66$, (see Figs.~3b,c), which corresponds to a reduction of -1~dB below the vacuum fluctuations for the squeezed quadrature.

\clearpage

\end{document}